\documentclass[longauth]{aa}

\usepackage{graphicx}
\usepackage[dvipsnames]{xcolor}
\usepackage[T1]{fontenc}
\usepackage[utf8]{inputenc}
\usepackage{lmodern}
\usepackage{txfonts}

\usepackage[colorlinks,bookmarks]{hyperref}
\definecolor{linkblue}{rgb}{0,0,0.8}
\definecolor{linkgreen}{rgb}{0,0.5,0}
\hypersetup{linkcolor=linkblue, citecolor=linkgreen, urlcolor=linkblue}

\usepackage{euclid}
\usepackage{bm}
\usepackage[normalem]{ulem}
\def\be{\begin{equation}}
\def\ee{\end{equation}}
\def\ba{\begin{eqnarray}}
\def\ea{\end{eqnarray}}
\def\eqi{\begin{equation}}
\def\eqf{\end{equation}}
\def\eqia{\begin{eqnarray}}
\def\eqfa{\end{eqnarray}}
\def\lcdm{$\Lambda$CDM\xspace}

\def\lltb{$\Lambda$LTB\xspace}
\def\nn{\nonumber}

\usepackage[nameinlink,noabbrev]{cleveref}
\Crefname{equation}{Eq.}{Eqs.}
\Crefname{eqnarray}{Eq.}{Eqs.}
\Crefname{section}{Sect.}{Sects.}
\Crefname{figure}{Fig.}{Figs.}
\crefname{equation}{Equation}{Equations}
\crefname{section}{Section}{Sections}
\crefname{figure}{Figure}{Figures}

\defcitealias{IST:paper1}{EC20}
\usepackage{amstext}

\begin{document}

\title{\Euclid: Forecast constraints on consistency tests of the $\Lambda$CDM model
\thanks{This paper is published on behalf of the Euclid Consortium.}}

\titlerunning{\Euclid: Forecast constraints on consistency tests of the $\Lambda$CDM model}

\author{S.~Nesseris$^{1}$\thanks{\email{savvas.nesseris@uam.es}}, D.~Sapone$^{2}$, M.~Martinelli$^{1}$, D.~Camarena$^{3}$, V.~Marra$^{4,5,6}$, Z.~Sakr$^{7,8}$, J.~Garcia-Bellido$^{9}$, C.J.A.P.~Martins$^{10,11}$, C.~Clarkson$^{12}$, A.~Da Silva$^{13,14}$
% ...
}

\institute{$^{1}$ Instituto de F\'isica Te\'orica UAM-CSIC, Campus de Cantoblanco, E-28049 Madrid, Spain\\
$^{2}$ Departamento de F\'isica, FCFM, Universidad de Chile, Blanco Encalada 2008, Santiago, Chile\\
$^{3}$ PPGCosmo, Universidade Federal do Esp\'{i}rito Santo, 29075-910,Vit\'{o}ria, ES, Brazil\\
$^{4}$ N\'{u}cleo Cosmo-ufes \& Departamento de F\'{i}sica, Universidade Federal do Esp\'{i}rito Santo, 29075-910, Vit\'{o}ria, ES, Brazil\\
$^{5}$ IFPU, Institute for Fundamental Physics of the Universe, via Beirut 2, 34151 Trieste, Italy\\
$^{6}$ INAF-Osservatorio Astronomico di Trieste, Via G. B. Tiepolo 11, I-34131 Trieste, Italy\\
$^{7}$ Universit\'e St Joseph; Faculty of Sciences, Beirut, Lebanon\\
$^{8}$ Institut de Recherche en Astrophysique et Plan\'etologie (IRAP), Universit\'e de Toulouse, CNRS, UPS, CNES, 14 Av. Edouard Belin, F-31400 Toulouse, France\\
$^{9}$ Departamento de F\'isica Te\'orica, Facultad de Ciencias, Universidad Aut\'onoma de Madrid, 28049 Cantoblanco, Madrid, Spain\\
$^{10}$ Centro de Astrof\'{\i}sica da Universidade do Porto, Rua das Estrelas, 4150-762 Porto, Portugal\\
$^{11}$ Instituto de Astrof\'isica e Ci\^encias do Espa\c{c}o, Universidade do Porto, CAUP, Rua das Estrelas, PT4150-762 Porto, Portugal\\
$^{12}$ School of Physics and Astronomy, Queen Mary University of London, Mile End Road, London E1 4NS, UK\\
$^{13}$ Departamento de F\'isica, Faculdade de Ci\^encias, Universidade de Lisboa, Edif\'icio C8, Campo Grande, PT1749-016 Lisboa, Portugal\\
$^{14}$ Instituto de Astrof\'isica e Ci\^encias do Espa\c{c}o, Faculdade de Ci\^encias, Universidade de Lisboa, Campo Grande, PT-1749-016 Lisboa, Portugal\\
$^{15}$ Universit\'e de Gen\`eve, D\'epartement de Physique Th\'eorique and Centre for Astroparticle Physics, 24 quai Ernest-Ansermet, CH-1211 Gen\`eve 4, Switzerland\\
$^{16}$ AIM, CEA, CNRS, Universit\'{e} Paris-Saclay, Universit\'{e} de Paris, F-91191 Gif-sur-Yvette, France\\
$^{17}$ Institute of Space Sciences (ICE, CSIC), Campus UAB, Carrer de Can Magrans, s/n, 08193 Barcelona, Spain\\
$^{18}$ Institut d’Estudis Espacials de Catalunya (IEEC), Carrer Gran Capit\'a 2-4, 08034 Barcelona, Spain\\
$^{19}$ Institute of Cosmology and Gravitation, University of Portsmouth, Portsmouth PO1 3FX, UK\\
$^{20}$ INAF-Osservatorio di Astrofisica e Scienza dello Spazio di Bologna, Via Piero Gobetti 93/3, I-40129 Bologna, Italy\\
$^{21}$ Max Planck Institute for Extraterrestrial Physics, Giessenbachstr. 1, D-85748 Garching, Germany\\
$^{22}$ INAF-Osservatorio Astrofisico di Torino, Via Osservatorio 20, I-10025 Pino Torinese (TO), Italy\\
$^{23}$ INFN-Sezione di Roma Tre, Via della Vasca Navale 84, I-00146, Roma, Italy\\
$^{24}$ Department of Mathematics and Physics, Roma Tre University, Via della Vasca Navale 84, I-00146 Rome, Italy\\
$^{25}$ INAF-Osservatorio Astronomico di Capodimonte, Via Moiariello 16, I-80131 Napoli, Italy\\
$^{26}$ INAF-IASF Milano, Via Alfonso Corti 12, I-20133 Milano, Italy\\
$^{27}$ Port d'Informaci\'{o} Cient\'{i}fica, Campus UAB, C. Albareda s/n, 08193 Bellaterra (Barcelona), Spain\\
$^{28}$ Institut de F\'{i}sica d’Altes Energies (IFAE), The Barcelona Institute of Science and Technology, Campus UAB, 08193 Bellaterra (Barcelona), Spain\\
$^{29}$ INAF-Osservatorio Astronomico di Roma, Via Frascati 33, I-00078 Monteporzio Catone, Italy\\
$^{30}$ Department of Physics "E. Pancini", University Federico II, Via Cinthia 6, I-80126, Napoli, Italy\\
$^{31}$ INFN section of Naples, Via Cinthia 6, I-80126, Napoli, Italy\\
$^{32}$ Dipartimento di Fisica e Astronomia ''Augusto Righi'' - Alma Mater Studiorum Universit\'a di Bologna, Viale Berti Pichat 6/2, I-40127 Bologna, Italy\\
$^{33}$ INAF-Osservatorio Astrofisico di Arcetri, Largo E. Fermi 5, I-50125, Firenze, Italy\\
$^{34}$ Institut national de physique nucl\'eaire et de physique des particules, 3 rue Michel-Ange, 75794 Paris C\'edex 16, France\\
$^{35}$ Centre National d'Etudes Spatiales, Toulouse, France\\
$^{36}$ Institute for Astronomy, University of Edinburgh, Royal Observatory, Blackford Hill, Edinburgh EH9 3HJ, UK\\
$^{37}$ European Space Agency/ESRIN, Largo Galileo Galilei 1, 00044 Frascati, Roma, Italy\\
$^{38}$ ESAC/ESA, Camino Bajo del Castillo, s/n., Urb. Villafranca del Castillo, 28692 Villanueva de la Ca\~nada, Madrid, Spain\\
$^{39}$ Univ Lyon, Univ Claude Bernard Lyon 1, CNRS/IN2P3, IP2I Lyon, UMR 5822, F-69622, Villeurbanne, France\\
$^{40}$ Institute of Physics, Laboratory of Astrophysics, Ecole Polytechnique F\'{e}d\'{e}rale de Lausanne (EPFL), Observatoire de Sauverny, 1290 Versoix, Switzerland\\
$^{41}$ Mullard Space Science Laboratory, University College London, Holmbury St Mary, Dorking, Surrey RH5 6NT, UK\\
$^{42}$ Department of Astronomy, University of Geneva, ch. d\'Ecogia 16, CH-1290 Versoix, Switzerland\\
$^{43}$ Universit\'e Paris-Saclay, CNRS, Institut d'astrophysique spatiale, 91405, Orsay, France\\
$^{44}$ Department of Physics, Oxford University, Keble Road, Oxford OX1 3RH, UK\\
$^{45}$ INFN-Padova, Via Marzolo 8, I-35131 Padova, Italy\\
$^{46}$ Istituto Nazionale di Astrofisica (INAF) - Osservatorio di Astrofisica e Scienza dello Spazio (OAS), Via Gobetti 93/3, I-40127 Bologna, Italy\\
$^{47}$ Istituto Nazionale di Fisica Nucleare, Sezione di Bologna, Via Irnerio 46, I-40126 Bologna, Italy\\
$^{48}$ INAF-Osservatorio Astronomico di Padova, Via dell'Osservatorio 5, I-35122 Padova, Italy\\
$^{49}$ Universit\"ats-Sternwarte M\"unchen, Fakult\"at f\"ur Physik, Ludwig-Maximilians-Universit\"at M\"unchen, Scheinerstrasse 1, 81679 M\"unchen, Germany\\
$^{50}$ Institute of Theoretical Astrophysics, University of Oslo, P.O. Box 1029 Blindern, N-0315 Oslo, Norway\\
$^{51}$ Jet Propulsion Laboratory, California Institute of Technology, 4800 Oak Grove Drive, Pasadena, CA, 91109, USA\\
$^{52}$ von Hoerner \& Sulger GmbH, Schlo{\ss}Platz 8, D-68723 Schwetzingen, Germany\\
$^{53}$ Max-Planck-Institut f\"ur Astronomie, K\"onigstuhl 17, D-69117 Heidelberg, Germany\\
$^{54}$ Aix-Marseille Univ, CNRS/IN2P3, CPPM, Marseille, France\\
$^{55}$ Department of Physics and Helsinki Institute of Physics, Gustaf H\"allstr\"omin katu 2, 00014 University of Helsinki, Finland\\
$^{56}$ NOVA optical infrared instrumentation group at ASTRON, Oude Hoogeveensedijk 4, 7991PD, Dwingeloo, The Netherlands\\
$^{57}$ Argelander-Institut f\"ur Astronomie, Universit\"at Bonn, Auf dem H\"ugel 71, 53121 Bonn, Germany\\
$^{58}$ Dipartimento di Fisica e Astronomia “Augusto Righi” - Alma Mater Studiorum Università di Bologna, via Piero Gobetti 93/2, I-40129 Bologna, Italy\\
$^{59}$ INFN-Sezione di Bologna, Viale Berti Pichat 6/2, I-40127 Bologna, Italy\\
$^{60}$ Institute for Computational Cosmology, Department of Physics, Durham University, South Road, Durham, DH1 3LE, UK\\
$^{61}$ California institute of Technology, 1200 E California Blvd, Pasadena, CA 91125, USA\\
$^{62}$ INFN-Bologna, Via Irnerio 46, I-40126 Bologna, Italy\\
$^{63}$ Observatoire de Sauverny, Ecole Polytechnique F\'ed\'erale de Lau- sanne, CH-1290 Versoix, Switzerland\\
$^{64}$ European Space Agency/ESTEC, Keplerlaan 1, 2201 AZ Noordwijk, The Netherlands\\
$^{65}$ Department of Physics and Astronomy, University of Aarhus, Ny Munkegade 120, DK–8000 Aarhus C, Denmark\\
$^{66}$ Perimeter Institute for Theoretical Physics, Waterloo, Ontario N2L 2Y5, Canada\\
$^{67}$ Department of Physics and Astronomy, University of Waterloo, Waterloo, Ontario N2L 3G1, Canada\\
$^{68}$ Centre for Astrophysics, University of Waterloo, Waterloo, Ontario N2L 3G1, Canada\\
$^{69}$ Institute of Space Science, Bucharest, Ro-077125, Romania\\
$^{70}$ Dipartimento di Fisica e Astronomia “G.Galilei", Universit\'a di Padova, Via Marzolo 8, I-35131 Padova, Italy\\
$^{71}$ Centro de Investigaciones Energ\'eticas, Medioambientales y Tecnol\'ogicas (CIEMAT), Avenida Complutense 40, 28040 Madrid, Spain\\
$^{72}$ Instituto de Astrof\'isica e Ci\^encias do Espa\c{c}o, Faculdade de Ci\^encias, Universidade de Lisboa, Tapada da Ajuda, PT-1349-018 Lisboa, Portugal\\
$^{73}$ Universidad Polit\'ecnica de Cartagena, Departamento de Electr\'onica y Tecnolog\'ia de Computadoras, 30202 Cartagena, Spain\\
$^{74}$ Kapteyn Astronomical Institute, University of Groningen, PO Box 800, 9700 AV Groningen, The Netherlands\\
$^{75}$ Infrared Processing and Analysis Center, California Institute of Technology, Pasadena, CA 91125, USA\\
$^{76}$ INAF-Osservatorio Astronomico di Brera, Via Brera 28, I-20122 Milano, Italy\\
$^{77}$ INFN-Sezione di Torino, Via P. Giuria 1, I-10125 Torino, Italy\\
$^{78}$ Dipartimento di Fisica, Universit\'a degli Studi di Torino, Via P. Giuria 1, I-10125 Torino, Italy\\
$^{79}$ Universit\'e de Paris, CNRS, Astroparticule et Cosmologie, F-75013 Paris, France\\
}

\authorrunning{S.~Nesseris et al.}


\abstract{
%
%
{\bf Context}: The standard cosmological model is based on the fundamental assumptions of a spatially homogeneous and isotropic universe on large scales. An observational detection of a violation of these assumptions at any redshift would immediately indicate the presence of new physics.

{\bf Aims}: We quantify the ability of the \Euclid mission, together with contemporary surveys, to improve the current sensitivity of null tests of the canonical cosmological constant $\Lambda$ and the cold dark matter ($\Lambda$CDM) model in the redshift range $0<z<1.8$.

{\bf Methods:}
We considered both currently available data and simulated \Euclid and external data products based on a $\Lambda$CDM fiducial model, an evolving dark energy model assuming the Chevallier-Polarski-Linder (CPL) parameterization or an inhomogeneous Lema\^{\i}tre-Tolman-Bondi model with a cosmological constant $\Lambda$ ($\Lambda$LTB), and carried out two separate but complementary analyses: a machine learning reconstruction of the null tests based on genetic algorithms, and a theory-agnostic parametric approach based on Taylor expansion and binning of the data, in order to  avoid assumptions about any particular model.

{\bf Results:}
We find that in combination with external probes, \Euclid can improve current constraints on null tests of the $\Lambda$CDM by approximately a factor of three when using the machine learning approach and by a further factor of two in the case of the parametric approach. However, we also find that in certain cases, the parametric approach may be biased against or missing some features of models far from $\Lambda$CDM.

{\bf Conclusions:}
Our analysis highlights the importance of synergies between \Euclid and other surveys. These synergies are crucial for providing tighter constraints over an extended redshift range for a plethora of different consistency tests of some of the main assumptions of the current cosmological paradigm.
}


\maketitle
%

\section{Introduction \label{sec:intro}}

Modern cosmology has been built on a combination of theoretical considerations, observations, and simulations~\citep{2020coce.book.....P}. Illustrative of the last two is the cosmological principle, which combines the observation that there seems to be no preferred direction in the sky (local isotropy), with the nonempirical assessment that we do not occupy a special place in the cosmos (Copernican principle). Together, they imply that the Universe on the largest scales must be spatially homogeneous (no special place)  and isotropic (no special direction), thereby motivating its description by the  Friedmann-Lema\^{\i}tre-Robertson-Walker (FLRW) geometry.

In addition to symmetries, dynamical prescriptions must be specified in order to model the history of cosmic expansion within the FLRW model. This aspect has mostly come from theoretical physics and observation. In standard lore, the coupling of matter and space-time geometry is described by the hitherto unchallenged theory of general relativity. At late times, which is the focus of the present article, the energy budget of the Universe is largely dominated by pressureless matter. The recent acceleration of cosmic expansion is attributed to Einstein's cosmological constant~$\Lambda$, which may also be interpreted as a negative-pressure fluid with an equation of state~$w=p/\rho=-1$; and the curvature of homogeneity hypersurfaces vanishes.

The set of all these features is collectively referred to as the standard cosmological model, or \lcdm. The \lcdm also includes initial conditions set by the inflationary paradigm. To date, the \lcdm is practically compatible with all cosmological observations~\citep{Aghanim:2018eyx}, and hence so are its underlying assumptions. Although it is intensely discussed in the community, the so-called $H_0$ tension~\citep{DiValentino:2021izs}, the $\sigma_8$ tension~\citep{Sakr:2018new}, and others~\citep{Perivolaropoulos:2021jda} have not yet proved convincing enough to claim the discovery of new physics.

Next-generation surveys, such as the \Euclid mission, will represent a leap forward in testing the \lcdm and its assumptions, for the amount and quality of their data will be unprecedented. The goal of this article is to assess the performance of a number of null tests for \lcdm in the context of \Euclid. These tests are designed to detect any deviation from the history of cosmic expansion as predicted by the concordance model. They are agnostic in the sense that they do not require any specific alternative to be tested against the \lcdm; they merely address the question whether \Euclid has detected new physics. Nevertheless, if one or several of the tests reviewed here were to reject the \lcdm, their variety would help identify which underlying assumption of the \lcdm would be violated: Is the Universe inhomogeneous or anisotropic on large scales? Is dark energy different from $\Lambda$? Do dark matter and dark energy interact? Is the Universe spatially curved? And so on.

Deviations from spatial homogeneity have previously been constrained with a plethora of different probes;~see \cite{Clarkson:2012bg} for a review. Examples of probes are standard candles \citep{Chiang:2017yrq}, the fossil record of galaxies \citep{Heavens:2011mr}, the BOSS DR12 quasar sample \citep{Laurent:2016eqo}, both spectroscopic \citep{Scrimgeour:2012wt} and  photometric redshift surveys  \citep{Alonso:2014xca}, the kinetic Sunyaev-Zeldovich (kSZ) effect on the cosmic microwave background (CMB)  \citep{Reichardt:2020jrr}, and measurements of our peculiar velocity \citep{Nadolny:2021hti}. On the theory side, several tests have been proposed, for example, by considering the propagation of light rays in an inhomogeneous universe using numerical relativity \citep{Giblin:2016mjp}, using distances to directly constrain the spatial curvature of the Universe \citep{Clarkson:2007pz,Clarkson:2012bg,Valkenburg:2012td}, and using inhomogeneous Lema\^{\i}tre-Tolman-Bondi (LTB) models  \citep{GarciaBellido:2008nz,February:2009pv,Redlich:2014gga,Valkenburg:2012td,Camarena:2021mjr}, and consistency tests have been conducted based on dynamical probes such as growth rate data \citep{Nesseris:2014mfa,Nesseris:2014qca} and through a linear model formalism \citep{Marra:2017pst}.

Future surveys are mainly expected to provide stringent constraints on null tests of $\Lambda$CDM via very accurate and precise distance measurements of baryon acoustic oscillations (BAO), type Ia supernovae (SNe), and lensing shear correlations. In this work we explicitly focus on \Euclid, which is an M-class space mission of the European Space Agency (ESA) scheduled for launch in 2022 \citep{Racca:2016qpi}. The near-infrared spectrophotometric instrument \citep{NISP_paper} and the visible imager \citep{VIS_paper} are carried on board the spacecraft. These instruments will jointly perform a spectroscopic and a photometric galaxy survey over 15\,000 deg$^2$ of sky, aiming to map the geometry of the Universe and measure the growth of structures up to $z\sim 2$~\citep{Laureijs:2011gra}.

The main cosmological probes of \Euclid will be galaxy clustering from the spectroscopic survey, and galaxy clustering and weak lensing from the photometric survey. Spectroscopic galaxy surveys offer much higher radial precision, but for fewer objects, while photometric surveys target a larger number of galaxies, but the redshift uncertainties are also larger. In particular, given its high spectroscopic accuracy, \textit{Euclid} will have very precise galaxy clustering constraints that also include the radial (i.e., along the line of sight) dimension. We here create mock BAO data using \Euclid spectroscopic survey specifications following the Fisher matrix approach as in \citet[][]{IST:paper1}, hereafter EC20.

We also highlight some of the synergies between \Euclid and other large-scale structure surveys, namely that of the Dark Energy Spectroscopic Instrument (DESI) \citep{DESI2016} and the Legacy Survey of Space and Time (LSST) survey, performed at the Vera C. Rubin Observatory \citep{Abell:2009aa}. Both will probe the expansion history of the Universe and its large-scale structure (LSS) while being complementary to \Euclid redshift ranges, thus greatly extending the redshift range of our constraints.

\citet{ReviewDoc} presented forecast constraints on deviations from spatial homogeneity and the Copernican principle by performing a joint analysis of the \Euclid galaxy survey \citep{Laureijs:2011gra} and a stage IV supernova mission \citep[][assuming SNAP as a concrete example]{DEtaskforce}. In this work we update these results by relying on more recent \Euclid specifications \citepalias[see][]{IST:paper1}, we  explore synergies with other surveys, and we extend the variety of tests. Moreover, we also implement machine learning and other model-independent approaches to reconstruct these tests to avoid theoretical biases.

The outline of our paper is as follows: in \Cref{sec:theory} we review the theoretical background of the fundamental assumptions of the standard cosmological model and various ways in which they can be broken. In \Cref{sec:tests} we summarize the null tests we used in our analysis. In \Cref{sec:cur_data} we describe the analysis methods we followed to test the fundamental assumptions of the standard cosmological model, using the currently available data and a machine learning approach. In \Cref{sec:fore_data} we describe how we produced mock data based on three different cosmologies: the vanilla
\lcdm model, which we used as our null hypothesis; an evolving dark energy equation of state $w(a)$ based on the Chevallier-Polarski-Linder (CPL) parameterization \citep{Chevallier:2000qy,Linder:2002et}, in order to examine the response of our null tests to different expansion histories; and a fiducial cosmology based on the LTB metric. The results of our analysis are then  discussed in \Cref{sec:currentresults} for the currently available data and in \Cref{sec:fore_res} for the mock data. We finally draw our conclusions in \Cref{sec:conclusions}.

\section{Theoretical background \label{sec:theory}}

The consistency tests to be considered in this work are motivated by various alternatives to the $\Lambda$CDM model that have been proposed in the literature. We consider deviations from this baseline model that range from simple modifications of the expansion history to models whose common feature is the relaxation of either the assumption of isotropy or that of spatial homogeneity of the Universe, or both.

\paragraph{CPL.}
As an example of the first category concerning modifications of the expansion history, we focus on the CPL parameterization of dark energy (DE), where the equation-of-state parameter $w$ for this component deviates from the constant $w=-1$ value that it takes in $\Lambda$CDM and is allowed to vary with redshift. In this parameterization, the DE equation-of-state parameter is given by \citep{Chevallier:2000qy,Linder:2002et}
\be\label{eq:CPLpar}
w(z) = w_0 + w_a \, \frac{z}{1+z}\, ,
\ee
where $w_0$ and $w_a$ are free parameters.

In the second category, which contains models that relax the assumption of isotropy or spatial homogeneity of the Universe, possible alternatives for our purposes may be grouped into five phenomenological approaches to modify the hypotheses described above.

\paragraph{Dipole.}
The first phenomenological approach is a simple deviation from the Copernican principle, whose effect is nonperturbative. Its origin might be the presence of a strong attractor affecting the observer's position. A typical observer will measure a high  dipole or other multipoles in the cosmic microwave background (CMB) spectrum due to the change in its peculiar velocity with respect to the CMB frame \citep[see][]{Naselsky:2011jp}, but also due to changes in the gradients of the expansion rate and the divergence of the shear constructed from the two now-measured directional Hubble parameters; see \citet{Valkenburg:2012td}.

\paragraph{LTB.}
Next we mention the class of LTB models proposed by \citet{Tomita:1999qn}, \citet{Celerier:1999hp} and \citet{Alnes:2005rw} as an alternative to DE \citep[see also][]{Moffat:1994qy,Mustapha:1998jb}.  This is an inhomogeneous universe model with spherical symmetry around a point. It therefore exhibits local isotropy at that point only. The possibility of a central gigaparsec-scale void  could explain SNe observations without requiring either an exotic type of repulsive content or any modification of gravity; see  \citet[]{GarciaBellido:2008nz}. The LTB metric is given by
\begin{equation}
\mathrm{d}s^{2}=-c^2\mathrm{d}t^{2}+\frac{\left[R'(t,r)\right]^{2}}
{1-K(r)}\mathrm{d}r^{2}+R^{2}(t,r)\mathrm{d}\Omega^{2}\,,
\label{eq:LTB}
\end{equation}
where $K(r)$ is an arbitrary function; a prime denotes a partial spatial derivative with respect to the radial dimension $r$. If $R=a(t)\,r$ and $K(r)=k_0 r^{2}$, we recover the familiar FLRW metric. For a review of these models, see \citet{Marra:2011ct} and references therein. It is worth stressing that void models as alternatives to dark energy have been ruled out. The kinetic Sunyaev-Zeldovich (kSZ) signal in models without decaying modes is too strong~\citep{GarciaBellido:2008gd,Zhang:2010fa,Moss:2011ze,Bull:2011wi}. On the other hand, models with sizeable decaying modes (which might have a weak kSZ signal) are ruled out because of $y$-distortions \citep{Zibin:2011ma}.  The only possibility of saving these void models would be choosing specific (fine-tuned) initial conditions, which would be contrived, or inhomogeneous radiation and baryon fraction~\citep{Clarkson:2010ej}. Consequently, in the following, we consider the so-called $\Lambda$LTB model, that is, the LTB model with a cosmological constant. This constant provides all or most of the dark energy, while we can phenomenologically constrain a small void contribution.

\paragraph{Bianchi models.}
A third class of models that instead abandons the isotropy hypothesis is the Bianchi models (see \citet[]{Ellis:1968vb,Ellis:1998ct}). These are spatially homogeneous models that are endowed with a group of isometries that transitively act on the spatial hypersurfaces. One slight modification is obtained in the type-I Bianchi models, which are characterized by a vanishing spatial curvature, and contain as a particular case the flat FLRW models. Their line element can be written as
\begin{equation}
\mathrm{d}s^{2} = - c^2\mathrm{d}t^2 + a_x^2(t) \mathrm{d}x^2 + a_y^2(t) \mathrm{d}y^2 + a_z^2(t) \mathrm{d}z^2 \,,
\label{metric}
\end{equation}
where at a given time $t$, the metric is spatially homogeneous (i.e., it does not depend on the spatial coordinates). In this metric, a sphere of comoving test particles experiences a change in volume with time, and unlike the FLRW case, is distorted into an ellipsoid. The sphere is characterized by the average expansion
\begin{equation}
    3H = H_x+H_y+H_z\; ,
\end{equation}
where $H_x$, $H_y$ , and $H_z$ is the Hubble expansion rate along the $x$, $y,$ and $z$ dimensions, respectively, while the rate of shape deformation is described by the shear $\sigma(i,j)$, which is constructed from the differences between the expansion rates $H_i-H_j$, with $(i,j=1,2,3)$. This can be completed by specifying the components of the spatial metric $h_{\mu\nu}$ at a single point and a unit time-like normal vector to the hypersurface $n^\mu$ such that the four-metric is specified by $g_{\mu\nu} = h_{\mu\nu} - n_{\mu}\,n_{\nu}$. Projecting the equations onto the $n^\mu$ direction then gives a generalized Friedmann equation,
\begin{equation}
  3H^2  = 8 \pi G \rho + \sigma^2 -  {}^{(3)}R/2 + \Lambda \, c^2  \,, \label{eq:friedmann}\end{equation}
where $G$ is the gravitational constant, $\rho$ is the 
energy density of the matter content, $\sigma^2 \equiv \sigma_{\mu \nu} \sigma^{\mu\nu}/2$ is the shear scalar, $\sigma_{\mu\nu}=\theta_{\mu\nu}-(\theta/3)\,h_{\mu\nu}$ is the shear tensor and $\theta_{\mu\nu}=h^\alpha{}_\mu\,h^\beta{}_\nu\,n_{\alpha\beta}$ is the expansion tensor, $\theta$ is its trace, and $h_{\mu\nu}=g_{\mu\nu}+n_\mu\,n_\nu$ for a time-like unit vector $n^\mu$. Finally, ${}^{(3)}R$ is the Ricci scalar of the three-curvature of the spatial hypersurfaces, and $\Lambda$ is the cosmological constant.

\paragraph{Backreaction.}
Another phenomenological approach stems from the idea that inhomogeneities in our Universe back-react and affect the average dynamics within our causal horizon such that the observed acceleration might be attributed to their influence. This means that if we would like to compute the impact of the inhomogeneities directly, without requiring a highly symmetric solution of Einstein’s equations, one ansatz used in back-reaction theory is to construct average quantities that follow equations similar to those of the traditional FLRW model. The averaged scale factor $a_{\cal D}$, with additional contributions, can then be written
\begin{align}
\label{averagedhamilton}
3 \frac{{\dot a}^2_{\cal D}}{a^2_{\cal D}} - 8\pi G \langle{\varrho}\rangle-\Lambda c^2 &= - \frac{\langle{R}_{\cal D}\rangle+{Q}_{\cal D} }{2} \,, \\
3\frac{{\ddot a}_{\cal D}}{a_{\cal D}} + 4\pi G \langle{\varrho}\rangle -\Lambda c^2&= {Q}_{\cal D}\,, \\
\dot{\langle{\varrho}\rangle}_{\cal D} + 3\frac{{\dot a}_{\cal D}}{a_{\cal D}} \langle{\varrho}_{\cal D}\rangle&=0\,,
\end{align}
where the scale factor has been replaced by an average factor. The new kinematic back-reaction source term
\begin{equation}
\label{backreactionterm}
{Q}_{\cal D} =
\frac{2}{3}\left\langle{\Big(\theta - \langle{\theta}\rangle\Big)^2 }\right\rangle -
2\left\langle{\sigma^2}\right\rangle \,
\end{equation}
arises as a result of the expansion rate $\theta$  and shear $\sigma$ fluctuations. For reviews of the underlying mathematical formalism, see \citet{Buchert:2007ik}, \citet{Kolb:2009rp} and \citet{Andersson:2011za} and references therein. There is a lack of consensus about whether back-reaction is a correct framework, however, and whether inhomogeneities can generate acceleration; see \citet{Clarkson:2011zq}, \citet{Green:2014aga}, \citet{Buchert:2015iva} and \citet{Kaiser:2017hqn}.

\paragraph{Swiss-cheese models.}
Finally, we briefly mention the Swiss-cheese model universe \citep[]{Einstein:1945id,Marra:2007pm}, where inhomogeneous LTB patches (the holes) are embedded in a flat FLRW background (the cheese) in a way that the dynamics of the holes is scale independent. In other words, small holes will evolve in the same way as large holes, and isotropy is therefore preserved \citep{Marra:2007pm,Biswas:2007gi}. One simple approach is to assume no shell crossing and to define the initial conditions for each shell at the same time, and the FLRW density is matched at the boundary of a hole.

We describe below the phenomenological formulations that serve as null tests to detect deviations from the hypothesis of the homogeneity and isotropy of the Universe. To do this, we have to assume a fiducial cosmology for a comparison. Even if this were enough for our trigger tests to suggest that some of the assumptions above need to be revised, we decided to also add a more subtle treatment in which the fiducial and null tests are considered in the framework of some of the above models. Therefore, we considered two well-studied and widely used models that were proposed as extensions and alternatives to $\Lambda$CDM: the CPL model described above, and the $\Lambda$LTB model we detail below, which has recently been shown to remain a viable alternative to $\Lambda$CDM \citep{Camarena:2021mjr}.

\section{Consistency tests \label{sec:tests}}
We now present the set of consistency tests of the \lcdm model that we reconstruct below, using mock data from \Euclid and other contemporary surveys. We explicitly use a variety of null tests of the \lcdm model and of the FLRW metric itself, as each one is sensitive to different probes in the redshift range covered by \Euclid. For example, while the Om statistic directly probes the Hubble expansion, the global shear test probes the spatial homogeneity of the Universe, and so on. It is necessary to focus on specific tests because to probe the assumptions of isotropy and homogeneity, we need to assume the FRLW metric, along with a model for interpreting the observations. One way to do this is by forecasting constraints on consistency tests of the \lcdm model, which are the tests we propose in this section.

\subsection{Om statistic\label{sec:omtest}}
 The Hubble parameter at late times, when we can safely neglect radiation, is given in a flat \lcdm universe by
\be
H^2(z)=H^2_0\left[\Omega_{\rm m,0}\,(1+z)^3+1-\Omega_{\rm m,0}\right]\,,\label{eq:Hlcdm}
\ee
where $H_0$ is the Hubble constant, $\Omega_{\rm m,0}\equiv \rho_\mathrm{m}/\rho_\mathrm{c}$ is the fractional matter density parameter, and $\rho_\mathrm{c}$ is the critical density for which the Universe is flat. Then, solving for $\Omega_{\rm m,0}$ , we can create the so-called
$\textrm{Om}_\mathrm{H}(z)$ quantity, defined as
\begin{align}
\textrm{Om}_\mathrm{H}(z) &\equiv \frac{h^2(z)-1}{(1+z)^3-1}
\longrightarrow \Omega_{\rm m,0} \quad \textrm{implies~ $\Lambda$CDM}\,,
\label{eq:omh1}
\end{align}
which has to be constant and equal to the matter energy density $\Omega_{m,0}$ only when \lcdm is the true model describing the evolution of the Universe and where we have defined the dimensionless Hubble parameter $h(z)\equiv H(z)/H_0$ (see \citet[]{Sahni:2008xx,Zunckel:2008ti}).

In the zero-redshift limit, however, this quantity is ill defined, which can limit the  precision and accuracy of this test. This might also introduce biases if the numerator and denominator uncertainties do not behave in the same way. In general, however, any deviations of \Cref{eq:omh1} from a constant and its \lcdm value given by \Cref{eq:Hlcdm} imply that \lcdm does not hold, regardless of the $\Omega_{\rm m,0}$ value. Reconstructions of the $\textrm{Om}_\mathrm{H}(z)$ test with earlier data were performed in \citet[]{Nesseris:2010ep}.

\subsection{Extensions of the Om statistic with curvature}
We also considered the extended $\textrm{Om}_\mathrm{H}(z)$ statistic by \citet{Seikel:2012cs} when curvature is present. In this case, we can solve simultaneously the \lcdm Friedmann equations for both the matter and the curvature parameters, and doing so, we find the two tests
\begin{align}
\mathcal{O}_{\rm m}(z) &\equiv  2\, \frac{ (1+z)\,[1-h^2(z)]+z\,(2+z)\,h(z)\,h'(z)}{z^2\,(1+z)\,(3+z)}\,, \label{Om-hz}\\
\mathcal{O}_{\rm K}(z) &\equiv \frac{ 3\,(1+z)^2\,[h^2(z) -1]- 2z\, (3+3z+z^2)\,h(z)\,h'(z)}{z^2\,(1+z)\,(3+z)}\,, \label{OK-hz}
\end{align}
where a prime is a derivative with respect to the redshift $z,$ and again we have
\begin{align}
\mathcal{O}_{\rm m}(z) &\longrightarrow  \Omega_{\rm m,0}
\quad\text{implies $\Lambda$CDM}, \\
\mathcal{O}_{\rm K}(z) &\longrightarrow  \Omega_{\rm k,0}
\quad \text{implies $\Lambda$CDM},
\end{align}
where both conditions have to hold simultaneously if the curved $\Lambda$CDM model is true.

The main advantage of these null tests compared to the  $\textrm{Om}_\mathrm{H}(z)$ is that we do not need to make any assumptions on the spatial curvature of the Universe. Other possibilities that also include information from the distances were explored in \citet{Yahya:2013xma}.

\subsection{$r_0$ test for interactions in the dark sector}
Recently, \citet[]{vonMarttens:2018bvz} (see also \citealt{vonMarttens:2020apn}) proposed a null test that is based on the ratio of cold dark matter (CDM) and DE energy densities, that is, $r(z) =\rho_\textrm{CDM}(z)/\rho_\textrm{DE}(z)$, which for the \lcdm model  is equal to $r(z)=r_0\,(1 +z)^3$, with $r_0=\Omega_{\rm c,0}/\Omega_{\rm DE,0}$.
Then the Friedmann equation for \lcdm can be rewritten as
\be
h^2(z) =\Omega_{\rm d,0}\,\frac{1+r_0\,(1+z)^3}{1+r_0}+\Omega_{\rm b,0}(1+z)^3\,, \label{eq:Hz2}
\ee
where $\Omega_{\rm d,0}=\Omega_{\rm c,0}+\Omega_{\rm DE,0}=1-\Omega_{\rm b,0}$. We assume flatness, and we ignore radiation as we only consider low-redshift data.

Solving \Cref{eq:Hz2} for $r_0$, we can define a null test for models with interactions in the dark sector as
\be
r_0=\frac{1-\Omega_{\rm b,0}+\Omega_{\rm b,0}(1+z)^3-h^2(z)}{h^2(z)-(1+z)^3}, \label{eq:r0null}
\ee
which for \lcdm has to be constant at all redshifts. This test is again ill defined at low redshifts, however, and also requires an external prior on the baryon density $\Omega_{\rm b,0}$. The natural choice would be to use the value obtained either from CMB data, for instance, \textit{Planck}, or from Big Bang nucleosynthesis (BBN), but the two currently do not agree \citep{Pitrou:2020etk}. We therefore use the \textit{Planck} 2018 best-fit value below.

Moreover, the tests of \Cref{eq:omh1} and \Cref{eq:r0null} are only degenerate in the case of $\Lambda$CDM, where we may show that indeed,
\be
r_0(z)=\frac{\textrm{Om}_\mathrm{H}(z)-\Omega_\mathrm{b,0}}{1-\textrm{Om}_\mathrm{H}(z)}\longrightarrow \text{constant},
\ee
which implies $\Lambda$CDM. Furthermore, if either of \Cref{eq:omh1} or \Cref{eq:r0null} is a constant, then so is the other. However, $r_0$ does depend on the value of $\Omega_\mathrm{b,0}$ , and in general, the relation between $r_0(z)$ and $\textrm{Om}_\mathrm{H}(z)$ might be more complicated for non-\lcdm models. Furthermore, as discussed in \citet[][]{vonMarttens:2018iav}, a nontrivial $r_0(z)$ can be mapped directly into the DM-DE coupling function.

Overall, the $r_0(z)$ test probes for interactions in the dark sector, which somewhat exceeds just testing for deviations from the \lcdm model. We did not consider fiducial cosmologies with DM-DE interactions here, however.

\subsection{Global shear}
We can also explore another test of the Copernican principle in the form of the normalized global shear $\Sigma\equiv (H_L-H_T)/(H_L+2H_T)$, where $H_L$ and $H_T$ are the longitudinal and transverse Hubble rates of the LTB metric \citep[see][]{GarciaBellido:2008yq} and the term $H_L+2H_T$ in the denominator is there in order to make the test dimensionless. Then, we can rewrite the normalized global shear using physical quantities such as angular diameter distances as a function of redshift as
\begin{align}
\Sigma(z)
&= \frac{\sqrt{1-kr^2(z)}-h(z)d'(z)}
{3h(z)\,\frac{H_0}{c}\,D_{\rm A}(z)+2\sqrt{1-kr^2(z)}-2h(z)d'(z)}
\nn \\[1mm]
&\simeq
\frac{1-h(z)d'(z)}{3h(z)\,\frac{H_0}{c}\,D_{\rm A}(z)+2-2h(z)d'(z)}
\,,
\label{eq:GlobalShear}
\end{align}
where the primes denote a derivative with respect to the redshift $z,$ and we have defined the dimensionless comoving distance,
\be
d(z)=(1+z)\;\frac{H_0}{c}\;D_{\rm A}(z)=\frac{1}{\sqrt{-\Omega_{\rm k,0}}}\sin{\left[
\sqrt{-\Omega_{\rm k,0}}\int_0^z{\frac{\mathrm{d}x}{h(x)}}\right]}\,.\label{eq:distances}
\ee
In a homogeneous universe described by the FLRW metric, the global shear given by \Cref{eq:GlobalShear} is equal to zero because the two expansion rates are the same, regardless of the curvature \citep[see][]{GarciaBellido:2008yq}.

\subsection{Distance null tests}
Applying the Lagrange inversion theorem to the dimensionless luminosity and angular diameter distances, $d_{\rm L}(z,\Omega_{\rm m,0})\equiv c^{-1}H_0\,D_{\rm L}(z,\Omega_{\rm m,0})$ and  $d_{\rm A}(z,\Omega_{\rm m,0})\equiv c^{-1}H_0\,D_{\rm A}(z,\Omega_{\rm m,0}),$ respectively, we can solve for $\Omega_{\rm m,0}$ similarly to the $\textrm{Om}_\mathrm{H}(z)$ test; see \citet[]{Arjona:2019fwb}. Restricting ourselves to late times, when DE dominates the other components, we may safely neglect radiation and neutrinos. Thus, the analytical expression of the dimensional luminosity distance for the \lcdm model, assuming a flat Universe, but neglecting radiation and neutrinos, is given by
\begin{multline}
D_{\rm L}(z,\Omega_{\rm m,0})
= c(1+z) \int_0^z\frac{{\rm d} x}{H(x)}
= \frac{c}{H_0}\frac{2(1+z)}{\sqrt{\Omega_{\rm m,0}}}
\\[2mm]
\times
\left\{_2F_1\left(\frac{1}{2},\frac{1}{6};\frac{7}{6};\frac{\Omega_{\rm m,0}-1}{\Omega_{\rm m,0}}\right)-
\frac{_2F_1\left[\frac{1}{2},\frac{1}{6};\frac{7}{6};\frac{\Omega_{\rm m,0}-1}{\Omega_{\rm m,0}(1+z)^{3}}\right]}{\sqrt{1+z}}\right\}
,
\label{eq:dl}
\end{multline}
where $_2F_1\left(a,b;c;x\right)$ is a hypergeometric function.

To invert the previous equation, we first performed a series expansion on \Cref{eq:dl} around $\Omega_{\rm m,0}=1$ and kept the first $12$ terms in order to obtain a reliable unbiased estimation and avoid theoretical systematic uncertainties by having to truncate the series expansion. Alternatively, one may directly Taylor expand the integrand of \Cref{eq:dl} and then perform the integration term by term, so that the two approaches are equivalent.

Finally, we applied the Lagrange inversion theorem to actually invert the series and to write the matter density $\Omega_{\rm m,0}$ as a function of the dimensionless luminosity distance $d_{\rm L}$, that is, $\textrm{Om}_\textrm{dL}=\textrm{Om}_\textrm{dL}(z,d_{\rm L})$. For example, the first two terms of the  expansion are
\be
\textrm{Om}_\textrm{dL}(a,d_{\rm L})=1-\frac{7a\left(d_{\rm L}-\frac{2-2\sqrt{a}}{a}\right)}{6+\sqrt{a}\left(a^3-7\right)}+O(d^2_{\rm L})\,, \label{eq:OmdLnull}
\ee
where the scale factor $a$ is related to the redshift $z$ as $a=1/(1+z)$, and as mentioned, for the actual calculations, we kept the first $12$ terms in the expansion. As we show in the plots in later sections, this series expansion converges nicely in the redshift range of our data.

This null test has the main advantage that it does not require taking derivatives of the data as we use the luminosity distance directly. Similarly, if the distance duality relation (DDR) holds, as we do in this work, then the angular diameter distance can be calculated via
\begin{equation}
    D_{\rm A}(z)=\frac{D_{\rm L}(z)}{(1+z)^2}\,,
\end{equation}
and assuming the number conservation of photons, we can follow the same approach as for the luminosity distance and find through the Langrange inversion theorem an expression for the matter density  $\Omega_{\rm m,0}$ as a function of the dimensionless angular diameter distance $d_{\rm A}(z)$, that is,  $\textrm{Om}_\textrm{dA}=\textrm{Om}_\textrm{dA}(z,d_{\rm A})$. The first two terms of the expansion in this case are
\be
\textrm{Om}_\textrm{dA}(a,d_{\rm A})=1-\frac{7\left[d_{\rm A}+2a\left(\sqrt{a}-1\right)\right]}{a\left[6+\sqrt{a}\left(a^3-7\right)\right]}+O(d^2_{\rm A})\,. \label{eq:OmdAnull}
\ee
As mentioned earlier, we kept the first $12$ terms of the expansion for the actual calculations in this case as well.

\subsection{Curvature test}
A direct test of the Copernican principle is the curvature test of \citet{Clarkson:2007pz}, which considers the distance-redshift relation in an FLRW model of any curvature. Solving \Cref{eq:distances} for the curvature parameter, we find
\be
\label{eq:OK}
\Omega_{\rm k}(z)=\frac{\left[h(z)d'(z)\right]^2-1}{[d(z)]^2}\longrightarrow\text{constant}\,.
\ee
One way to interpret \Cref{eq:OK} is that it provides the means for measuring the current value of the curvature parameter today by using Hubble rate and distance data. In the context of the FLRW metric, this parameter is constant regardless of the dark energy model. Equivalently, we can also write the aforementioned condition as
\begin{multline}
   \label{C(z)}
\mathcal{C}(z)
= 1+h^2(z)\left\{d(z)\,d''(z)-\left[d'(z)\right]^2\right\}\\
+ h(z)h'(z)d(z)d'(z)
\longrightarrow 0\,.
\end{multline}
In general, for space times other than FLRW metrics (e.g., in LTB models), we have that $\mathcal{C}(z)\neq0$, or $\Omega_{\rm k}(z)\neq$\,const.

\begin{figure}[!t]
\centering
\includegraphics[width = 0.48\textwidth]{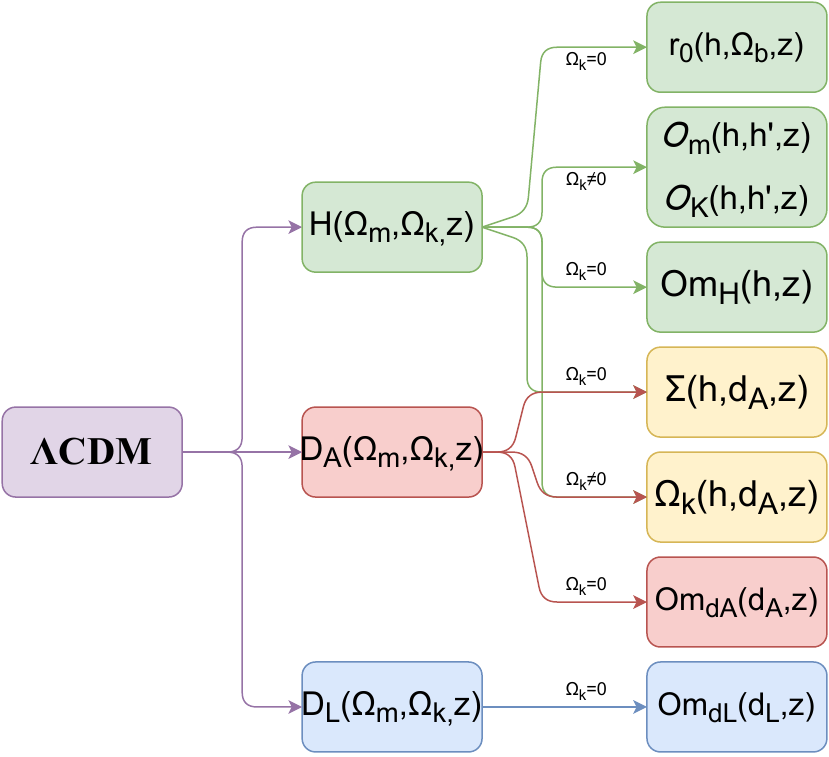}
\caption{Flowchart describing the consistency tests presented in \Cref{sec:tests}. The color-coding corresponds to green for the tests that require the Hubble parameter $H(z)$, red for those that require the angular diameter distance $D_\mathrm{A}(z)$, $\mathrm{yellow}=\mathrm{red} + \mathrm{green}$ for tests that require both $H(z)$ and $D_\mathrm{A}(z),$ and finally, blue for the test that requires the luminosity distance $D_\mathrm{L}(z)$. In all cases, we inverted or solved for crucial quantities (a process denoted by an arrow) that describe the expansion history of the Universe based on the \lcdm model, making assumptions about the curvature as required in each test (denoted above the arrows).
As an example, we can solve the Hubble parameter $H(\Omega_\mathrm{m,0},z)$ of \lcdm for $\Omega_\mathrm{m,0}$ , and assuming flatness, we can write the matter density parameter in terms of the dimensionless Hubble parameter $h(z)$, see \Cref{eq:omh1}. Clearly, $\mathrm{Om}_\mathrm{H}(z)\rightarrow \Omega_\mathrm{m,0}$ only if \lcdm is the correct description of the expansion of the Universe. A similar process is also followed for the remaining tests, which we can write in terms of the dimensionless distances $d_{\rm L}(z)$ and $d_{\rm A}(z)$.}
\label{fig:flow}
\end{figure}

As was shown in \citet[]{Sapone:2014nna}, however, the curvature test $\Omega_{\rm k}(z)$ diverges as $\sim1/z^2$ for low redshifts, hence we regularize the divergence, and instead consider the quantity $z^2 \Omega_{\rm k}(z)$ below. This is also similar to the complementary test of the Copernican principle that also uses the distances and the Hubble rate, as explored in \citet[]{Arjona:2021zac}.

Finally, in \Cref{fig:flow} we show a flowchart that systematically organizes the different tests we consider. We refer to \citet{Clarkson:2012bg} for a review.

\section{Analysis method of  currently available data\label{sec:cur_data}}

\subsection{Currently available data }
In order to constrain deviations from the \lcdm model and the FLRW metric, we now analyze a set of data providing information about the Hubble rate $H(z)$, the luminosity $D_{\rm L}(z),$ and the angular diameter distance $D_{\rm A}(z)$. Below, we describe the currently available BAO and SNe data and present our analysis method.

We first considered the BAO data, given via measurements of the ratio $d_z$, which is defined as
\be\label{eq:dz}
d_z\equiv \frac{r_{\rm s}(z_{\rm d})}{D_{\rm V}(z)}\,,
\ee
where $D_{\rm V}$ is the spherically averaged distance
\be
D_{\rm V}(z)=\left[(1+z)^2\, D_{\rm A}^2(z)\, \frac{c\,z}{H(z)}\right]^{1/3}\, ,
\ee
while $r_{\rm s}(z_{\rm d})$ is the comoving sound horizon at the drag epoch,
\be
r_{\rm s}(z_{\rm d})=\int_{z_{\rm d}}^\infty \frac{c_{\rm s}(z)}{H(z)} \; \text{d}z\, ,
\label{eq:sound-horizon-drag}
\ee
$c_{\rm s}(z)$ is the sound speed of the baryon-photon plasma, and $z_{\rm d}$ is the redshift at the drag epoch; see Eq.~(4) of \citet{Eisenstein:1997ik} for a fitting function that is accurate to about $2\%$ on average compared to numerical estimates of $r_{\rm s}(z_{\rm d})$ obtained via recombination codes. \citet[]{BOSS:2013rlg} and \citet[]{Aubourg:2014yra} obtained similar approximations for the sound horizon  that are accurate to about $0.1\%$, while \citet[]{Aizpuru:2021vhd} presented machine learning improved fitting functions, which are accurate to about $0.018\%$.

The actual BAO data considered here, described in terms of $d_z$ given by \Cref{eq:dz}, the Hubble distance $D_{\rm H}(z)=c/H(z),$ and the comoving angular diameter distance $D_{\rm M}(z)= (1+z)\, D_\mathrm{A}(z)$, are provided by the 
6dF galaxy survey (6dFGS) \citep{Beutler:2011hx}, the WiggleZ survey \citep{Blake:2012pj}, the third-year (Y3) data release of the dark energy survey (DES) \citep{DES:2021esc}, and the extended baryon oscillation spectroscopic survey (eBOSS) of the completed Sloan digital sky survey (SDSS-IV) \citep{eBOSS:2020yzd}. We refer to this combination of points as BAO and refer to Appendix A of \citet[]{Martinelli:2020hud} and  Table 3 of \citet{eBOSS:2020yzd} for their exact values and their likelihood.

We have considered that a fiducial cosmology is commonly adopted in order to convert the measured angular scales into distances for the BAO data; we corrected the BAO results to be consistent with our models where necessary. However, we did not model the change in the nonlinear effects that may modify and damp the position of the BAO. This could introduce systematic uncertainties of a few percent in the reconstructions \citep[see][]{Angulo:2007fw} and may lead to a lesser constraining power \citep[see, e.g.,][]{Anselmi:2017cuq}. The reason is that the BAO is essentially a feature at linear scales and the damping parameter is marginalized over. Moreover, modeling the nonlinear scales is nontrivial for theories beyond the \lcdm model and the FLRW metric and thus is beyond the scope of this paper. Therefore, we chose not to correct for these effects and left this for future and more focused work on the subject.

We also considered the SNe, which can constrain the luminosity distance $D_{\rm L}(z)$. The key observable is the apparent magnitude $m(z)$
\begin{equation}
    m(z) = M_0+5\log_{10}{\left[\frac{D_{\rm L}(z)}{\rm Mpc}\right]}+25\, ,
    \label{eq:lum_dist}
\end{equation}
where $M_0$ is the intrinsic magnitude of the supernova at a redshift $z$. Clearly, if no external information is provided, the Hubble constant $H_0$ is completely degenerate with $M_0$ as the SNe data alone cannot constrain these two quantities. Here we considered the updated Pantheon compilation of 1048 points from \citet{Scolnic:2017caz}, while for the likelihood, we used the expression given in  Appendix C of \citet{Conley:2011ku}, which is already marginalized over~$M_0$ and~$H_0$ and takes the covariances of the SNe data into account.

\subsection{Genetic algorithms\label{sec:GA}}

We now describe a machine learning approach, called genetic algorithms (GA). The GAs are a set of stochastic optimization methods that are commonly applied in the context of nonparametric reconstructions of data. They are inspired by the notion of grammatical evolution, and in particular, the genetic operations of mutation, that is, a random change in an individual, and crossover, that is, the combination of different individuals to form offspring. These operators behave as an environmental pressure, thus emulating the concept of natural selection in biology.

The reproductive success of a member of the population, or in other words, the probability that it will have offspring, is usually assumed to be proportional to its fitness, which is a measure of how accurately every member of the population fits the data. We quantified the fitness of the individuals via a standard $\chi^2$ statistic. Overall, the GA have been used to probe for extensions of the standard model \citep[]{Akrami:2009hp}; deviations from $\Lambda$CDM,
both at the background and the perturbation level in linear order \citep[]{Nesseris:2012tt,Arjona:2020kco,Arjona:2019fwb}; to reconstruct a variety of cosmological data, such as SNe data \citep[]{Bogdanos:2009ib, Arjona:2020doi}; or to perform reconstructions of null tests, for example, the so-called Om statistic and the curvature test \citep[]{Nesseris:2010ep, Nesseris:2013bia, Sapone:2014nna}.

We performed a simultaneous fit of the SNe and BAO data with the GA for the currently available data and for the mocks presented in later sections, via the following approach. During the initialization phase, a set of functions was randomly chosen among an orthogonal basis, in the classical sense of orthogonal polynomials for the related functions that include polynomials, exp, log, and other elementary functions, but also the  operators $+,-,\times,\div,\text{ and } \wedge$, where the wedge corresponds to exponentiation, for instance, for two functions $f(x)$ and $g(x)$, that would be $f\wedge g\equiv f^g=\exp\left[g*\log(f)\right]$, so that every member corresponds to a random guess for the Hubble rate $H(z)$ and the luminosity distance $D_\textrm{L}(z)$. Although we still assumed the validity of the distance duality relation between the luminosity and angular diameter distances, $D_\textrm{L}(z)=(1+z)^2\, D_\textrm{A}(z)$, as this case was tested in \citet{Martinelli:2020hud}, we relaxed the assumption of \Cref{eq:distances}, however, because if it is valid, then the global shear test $\Sigma(z)$ is automatically satisfied regardless of the reconstruction method (GA, etc). Thus, the GA evolves the three functions of redshift that correspond to the Hubble rate $H(z)$, the luminosity distance $D_\textrm{L}(z),$ and the angular diameter distance, given by $D_\textrm{A}(z)=D_\textrm{L}(z)/(1+z)^2$.

The GA does not a priori impose a prior on the functional space, and in principle, given a grammar, the output best-fitting functions are unconstrained in terms of their properties (other than they should fit the data very well). However, we do need to assume certain priors in order to ensure that the derived functions have physical meaning. The way this was done is twofold. First, we demanded that all functions were smooth, continuous, and differentiable across the redshift range covered by the data. This was done automatically in our implementation of the GA code. Second, we also assumed certain physical priors, for example, that the luminosity distance at z=0 was zero, to ensure that the resulting functions were realistic, but we made no assumption on a DE model.

After this initial population of function was set up, we then calculated the fitness of every member via a $\chi^2$ statistic, using the SNe and BAO data and their individual covariances simultaneously
as input, thus assessing the global fitness of the set of the three reconstructed functions. Next, using the  tournament selection approach (see \citet{Bogdanos:2009ib} for more details), a random set of the best-fitting functions in every generation was chosen, and the crossover and mutation operators were applied to the selected functions. Finally, we iterated this process thousands of times in order to ensure a good convergence of the algorithm, but we also performed runs with different random seeds, in order to avoid biasing the results by the choice of a specific random seed.

The final output of the GA after converged is a set of three continuous and differentiable functions of redshift for the Hubble parameter $H(z)$, the luminosity distance $D_\textrm{L}(z),$ and the angular diameter distance $D_\textrm{A}(z)$. In the case of the currently available data, we also numerically minimized the $\chi^2$ over the combination $r_{\rm s}(z_{\rm d}) h$, where $r_{\rm s}(z_{\rm d})$ is the comoving sound horizon at the drag epoch and $h=H_0/(100\ {\rm km\ s^{-1}\ Mpc^{-1}})$, in order to avoid any assumptions about the Hubble constant for the BAO physics at early times. However, this complication does not exist for the fiducial BAO data, as they have already been marginalized over that quantity and the angular diameter distance is directly available. Moreover, because the BAO data cannot independently constrain the combination $\Omega_{\rm b, 0}h^2$, we assumed the value $\Omega_{\rm b, 0} h^2=0.02225$ from Planck 2018 \citep{Aghanim:2018eyx} where necessary.

The uncertainties of the best-fit Hubble parameter $H(z)$ and the luminosity distance $D_\textrm{L}(z)$ were obtained from the GA via a path integral
approach developed by \citet{Nesseris:2012tt,Nesseris:2013bia}. Specifically, the uncertainties were calculated by functionally integrating the likelihood, that is, performing a path integral, over the whole functional space covered by the GA. This approach has been extensively tested by \citet{Nesseris:2012tt} and was found to agree with error estimates obtained via a bootstrap Monte Carlo method. We used the publicly available code \texttt{Genetic Algorithms} for the numerical implementation of the GA.\footnote{\url{https://github.com/snesseris/Genetic-Algorithms}}

\subsection{Binning and parameterized approach\label{sec:parapp}}

We can now describe our parameterized method for the null tests presented in the previous section to probe for deviations from \lcdm and spatial homogeneity. To do this, we followed a two-pronged approach.

First, we binned the Hubble rate and distance data in bins with an equal number of points, and then we calculated the various tests using the expressions in \Cref{sec:tests}. The first technique to explore the sensitivity of the tests we present consists of evaluating them in redshift bins. Each test is sensitive to a particular combination of the different observables and parameters, hence the binning technique had to be performed individually to take the proper dependence and the covariances into account. To compute the $\mathrm{Om}_{\rm H}$ and $r_0$ tests, we binned the $H(z)$ data at different redshifts and propagated their uncertainties to the final tests by taking the diagonal elements of the covariance matrices of the data. We verified following the approach of \citet[]{Nesseris:2014vra} that our choice of the covariance matrices did not affect the results by more than $1\%$.

Furthermore, to compute $\mathrm{Om}_{\rm dA}$ and $\mathrm{Om}_{\rm dL}$, we binned the corresponding distances. $D_{\rm A}(z)$ comes directly from the BAO mocks, whereas for $D_{\rm L}(z),$ we used the distance moduli of SNe by inverting \Cref{eq:lum_dist}. For the distance moduli, we further binned the data in 20 equally spaced redshift bins in order to have enough points and help compute the derivatives in a wider redshift range.

The uncertainties on the final $\mathrm{Om}_{\rm dL}$ test were obtained using a standard error propagation technique. Particular care was taken for the $\Sigma(z)$, $\Omega_{\rm k}(z)$, $\mathcal{O}_{\rm m}(z),$ and $\mathcal{O}_{\rm K}(z)$ tests. The global shear test $\Sigma(z)$ depends simultaneously on $H(z)$ and $D_{\rm A}(z)$ with its first derivative with respect to redshift, $D'_{\rm A}(z)$. The derivative of the angular diameter distance was obtained numerically considering two consecutive bins,
\be
D'_{\rm A}(z_i) = \frac{D_{\rm A}(z_{i+1})-D_{\rm A}(z_{i})}{z_{i+1}-z_i}\,.
\label{eq:num_der_da}
\ee
The uncertainties on the final test were obtained by constructing the Jacobian matrix, which takes  the derivatives of $\Sigma$ with respect to $H(z)$, $D_{\rm A}(z),$ and $D'_{\rm A}(z)$ into account and by then propagating the matrix with the covariance matrix of the data.

The curvature test $\Omega_{\rm k}(z)$ depends on $H(z)$, $d(z) = (1+z)\,c^{-1}H_0\,D_{\rm A}(z),$ and $d'(z)$, hence we recast the $\Omega_{\rm k}$ test as a function of the angular diameter distance in order to directly propagate the uncertainties from the data. We also adopted the numerical derivative for two consecutive bins for $d'(z)$, similar to \Cref{eq:num_der_da}, and the final uncertainties were obtained as for the $\Sigma(z)$ test.

The $\mathcal{O}_{\rm m}(z)$ and $\mathcal{O}_{\rm K}(z)$ terms only depend on $H(z)$ and its first derivative, $H'(z)$. The latter was evaluated numerically in the same way, and the uncertainties on the final tests were obtained using the propagation rule for both $H(z)$ and $H'(z)$. Finally, the number of bins for $\Sigma(z)$, $\Omega_{\rm k}(z)$, $\mathcal{O}_{\rm m}(z),$ and $\mathcal{O}_{\rm K}(z)$ is $n_{\rm data}-1$ because the test depends on the first derivatives of the observables.

Second, for each of the null tests we examined, we fit the binned data we derived in the previous step with a parametric form that is inspired by the CPL parameterization for the dark energy equation of state, shown in \Cref{eq:CPLpar}, in order to obtain forecast constraints from \Euclid for the deviations from the expected values and for the redshift trends of these null tests.

Specifically, each quantity we tested ($P$) was parameterized with the functional form
\be
P(z)=P_0+P_1 \frac{z}{1+z},\label{eq:CPL}
\ee
where $P_0$ is its value today and $P_1$ the first derivative, both evaluated at $a=1$. This simple parameterization, based on a Taylor expansion, allows us to characterize in a simple and model-independent fashion how \Euclid will be able to carry out the tests. The only exception to this form is the one chosen for the test of \Cref{eq:OK}, for which we multiplied the parameterization above by a factor $z^2$ in order to ensure that there is no divergence at the present time.

Using the binned values we obtained in the first step of this approach as input data, we fit the predictions given by \Cref{eq:CPL}, sampling the two free parameters through the publicly available sampler \texttt{Cobaya} \citep{Torrado:2020dgo}, which exploits the Metropolis-Hastings algorithm presented in \citet{Lewis:2002ah} and \citet{Lewis:2013hha}. In each case, we assumed flat priors on the two free parameters of the analysis. The results of this parameterized approach is denoted PA below.

We emphasize that the order of the expansion we used for the PA is somewhat arbitrary. We chose to truncate the expansion at first order, analogously with the CPL parameterization for the DE equation of state, but this choice has significant effects. Our results clearly show (see \Cref{sec:fore_res}), the PA at first order lacks the flexibility to accurately reconstruct several of the tests considered in this paper, especially when we do not consider the $\Lambda$CDM fiducial. The purpose of the PA here therefore is to perform a comparison with the GA, which clearly shows the trade-off between simplicity and accuracy of the reconstruction.

\begin{table}
\begin{center}
\setlength{\tabcolsep}{4pt}
\renewcommand{\arraystretch}{1.35}
\caption{Parameter values for the fiducial models we used for the mocks. The values for the \lcdm follow the fiducial of \citetalias{IST:paper1}; in particular, spatial flatness is assumed. $H_0$ is shown in units of km s$^{-1}$~Mpc$^{-1}$. \label{tab:fiducials}}
\begin{tabular}{lcccccccc}
\hline
\hline
model & $M_0$ & $\Omega_{\rm m,0}$ & $\Omega_{\rm b,0}h^2$ & $H_0$ & $w_0$ & $w_a$ & $\delta_0$ & $z_B$\\ \\
 \hline
\lcdm &$-19.3$ & $0.32$ & $0.02225$ & $67$ & $-1$ & $0$ & - & -\\
CPL& $-19.3$ & $0.32$ & $0.02225$ & $67$ & $-0.8$ & $-1$& - & - \\
\lltb &$-19.3$ & $0.32$ & $0.02225$ & $67$ & $-1$ & $0$&  $-0.65$ & $1.5$ \\
\hline
\hline
\end{tabular}\\
\end{center}
\end{table}

\section{Analysis method for the  mock data\label{sec:fore_data}}

\subsection{Fiducial cosmologies}

In order to forecast the ability of \Euclid to improve upon the sensitivity of the null tests mentioned in \Cref{sec:tests}, we relied on mock data based on a priori known fiducial cosmologies, which we briefly describe. In particular, we used the $\Lambda$CDM model, the CPL model ($w_0w_a$CDM), and the \lltb model to create simulated data sets for the SNe and the BAO measurements based on the \Euclid specifications.

First, for the fiducial cosmologies based on the \lcdm and CPL models, we used the values of the parameters shown in \Cref{tab:fiducials}. Both of these fiducial cosmologies assume no violation of the spatial homogeneity and isotropy of the FLRW metric, and the \lcdm fiducial was also used in \citetalias{IST:paper1}. In this case, we can jointly write the Hubble expansion rate at late times when we safely neglect radiation and assuming flatness ($\Omega_{\rm k,0}=0$) for the two models as
\be
\frac{H^2(a)}{H^2_0}=\Omega_{\rm m,0}\;a^{-3}+(1-\Omega_{\rm m,0})\; \exp\left[-3\int_1^a\frac{1+w(\alpha)}{\alpha}\;
{\rm d}\alpha\right]\,,
\label{eq:HlcdmDE}
\ee
where the dark energy equation of state for the CPL model follows \Cref{eq:CPLpar}, which includes the \lcdm model for $(w_0,w_{a})=(-1,0)$, while the scale factor is $a= 1/(1+z)$. For the CPL fiducial, we chose the parameters $(w_0,w_{a})=(-0.8,-1)$ as an extreme case, which corresponds to the combination  TT, TE, EE+lowE+lensing+SNe+BAO; see Fig.~30 of \citet[]{Aghanim:2018eyx}. Furthermore, we also assumed the values for the SNe absolute magnitude $M_0$, the baryon density $\Omega_{\rm b,0}h^2$ , and the Hubble rate today $H_0$ as given in \Cref{tab:fiducials}.

Regarding the \lltb mock, considering the LTB metric of \Cref{eq:LTB} and using Einstein's equations, we obtain the modified Friedmann equation, 
\be
H_\bot^2(r,t)
\equiv
\left[\frac{\dot{R}(r,t)}{R(r,t)}\right]^2 = \frac{2m(r)}{R^3(r,t)}+\frac{2 r^2k(r)\,M^2}{R^2(r,t)}+\frac{\Lambda\; c^2}{3}, \label{eq:LLTB_fried}
\ee
where the dot is a time derivative, and the radial dependent matter density
\be
\rho_{\rm m}(r,t)=\frac{m'(r)}{4\pi r^2 R'(r,t)R(r,t)^2},
\ee
where $M$ is an arbitrary mass scale, and $m(r)$ is the so-called Euclidean mass function.  We have conveniently recast $K(r) \equiv - 2 r^2 k(r) M^2$, where $k(r)$ is the curvature profile of the model.\footnote{Here, $H_\bot$ and $M$ have units of $\mathrm{Mpc}^{-1}$, $m$ has units of $\mathrm{Mpc,}$ and that $k(r)$ is dimensionless.} In addition to the mass function, $m(r)$, and the curvature profile, $k(r)$, the LTB solution introduces another free function: the so-called Big Bang function, $t_{\rm BB}(r)$, which appears as the constant of integration of \Cref{eq:LLTB_fried}, defined via
\be
t - t_{\rm BB}(r) = \int^{R(r,t)}_0 \frac{d x}{\sqrt{2m(r)x^{-1} + 2 r^2k(r)M^2 + \frac{\Lambda}{3} x^2}} .
\ee
We also assumed the compensated curvature profile,\begin{align}
k(r) &= k_{\rm b}+(k_{\rm c}-k_{\rm b})P_3(r/r_{\rm B})\,,\\
P_{n}(x) &=
\begin{cases}
1-\exp\left[-\frac{1}{x}(1-x)^n\right]
& 0 \le x<1, \\
0 & 1\le x,
\end{cases}
\end{align}
where $k_{\rm b}$ and $k_{\rm c}$ are the curvature outside and at the center of the spherical inhomogeneity, respectively, and $r_{\rm B}$ is the comoving radius of the inhomogeneity. In order to avoid the decaying modes on the matter density field, we set the Big Bang function to $t_{\rm BB} (r) = 0$.  The absence of decaying modes of the $\Lambda$LTB model ensures the agreement with the standard scenario of inflation. The remaining arbitrary function, $m(r)$, is effectively a gauge choice, which we fixed here by $m(r) = 4\pi M^2 r^3/3$ (see \citealt[][]{Biswas:2010xm} for more details).

\begin{figure}[!t]
\centering
\includegraphics[width = 0.44\textwidth]{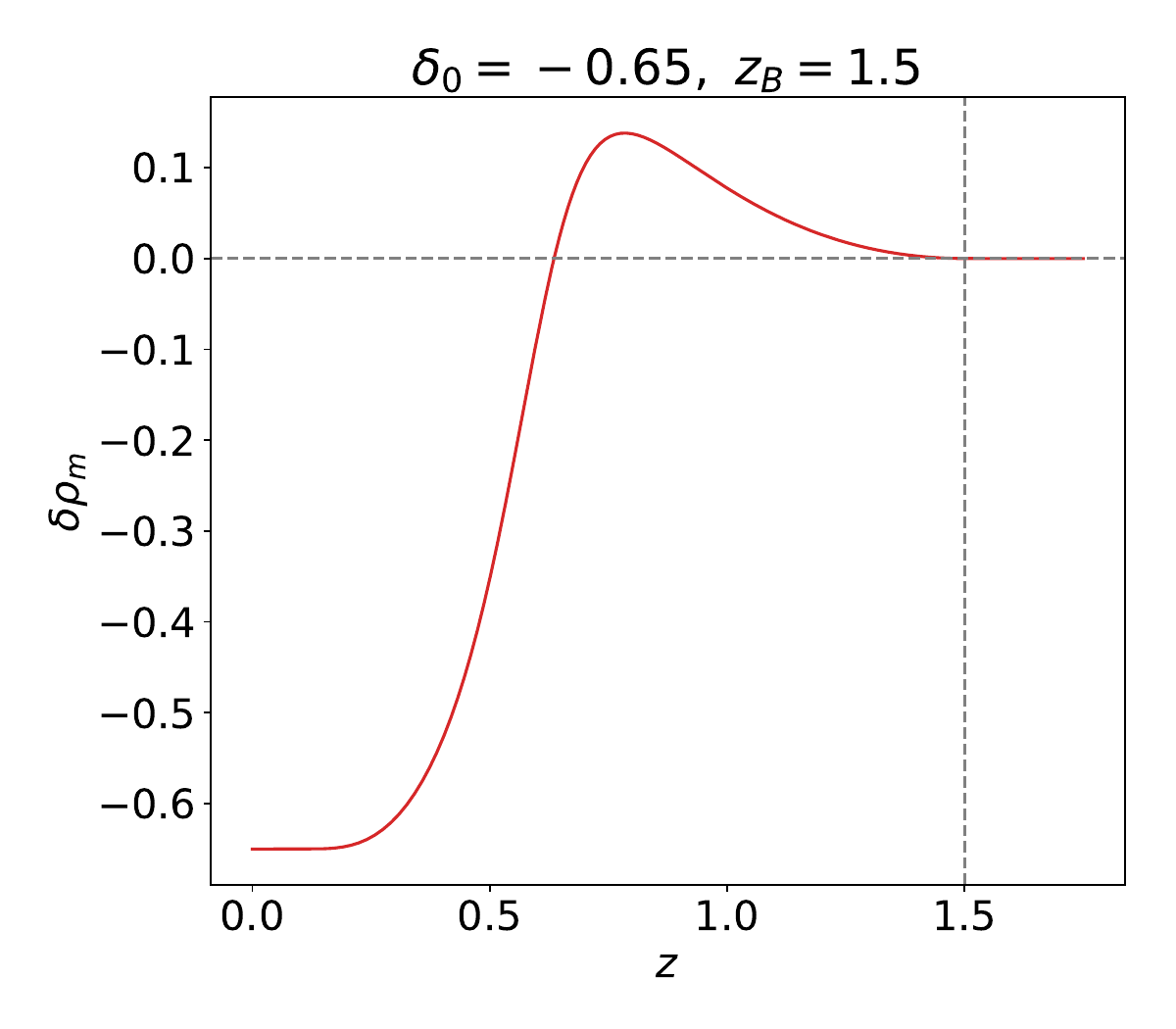}
\caption{Matter density contrast of the \lltb model as a function of redshift $z$ for $\delta_0=-0.65$ and $z_B=1.5$.}
\label{fig:LLTB}
\vspace*{0cm}
\end{figure}

Since we assumed a compensated model, the \lltb model becomes exactly \lcdm at scales $r>r_{\rm B}$. It is specified by the six usual \lcdm parameters plus the parameters introduced by the profile $k(r)$, namely $k_{\rm c}$ and $r_{\rm B}$ (from \Cref{eq:LLTB_fried}, note that $k_{\rm b} \equiv 4 \pi \Omega_{\rm k,0}/3 \Omega_{\rm m,0}$). Following \citet[]{Camarena:2021mjr}, we mapped $k_{\rm c}$ into the central density contrast $\delta_0 \equiv \rho_{\rm m}(0,t_0)/\rho_{\rm m}(r_{\rm B},t_0) -1$  and $r_{\rm B}$ into its corresponding redshift $z_{\rm B}$. We provide the relevant parameters used for the \lltb mocks in \Cref{tab:fiducials}, and in 
\Cref{fig:LLTB} we show the matter density contrast, defined as $\delta \rho_{\rm m} \equiv \rho_{\rm m}(r,t_0)/\rho_{\rm m}(r_{\rm B},t_0) -1$.
The values of the parameters $\delta_0$ and $z_{\rm B}$ were chosen for historical reasons. An underdensity of contrast $\approx -0.5$ up to a redshift of $\approx 0.7$ can indeed fit the luminosity-distance-redshift relation of the $\Lambda$CDM model, explaining away dark energy \citep[][]{Marra:2011ct}.

After establishing the three cosmologies that we wished to examine, we calculated the fiducial background quantities for the Hubble rate, the luminosity, and angular diameter distances as a function of redshift $z$. Finally, using the specifications of forthcoming surveys as described below, we created our mock BAO and SNe data.

While high precision is expected from forthcoming surveys such as \Euclid, it is crucial to ensure that they are accurate, and several efforts have been made to take observational systematic uncertainties into account that will affect the surveys. For example, in \citet{Paykari2020}, the observational systematic effects of the \Euclid VIS instrument were studied, also taking the modeling of the point spread function and the charge transfer inefficiency into account. While some of the specifications for future instruments might still change prior to completion, here we assumed that by the time the data arrive, the systematic effects described above will be well understood. Thus, we added all the relevant astrophysical systematic effects, such as the galaxy bias, as described in
the following subsections.

Finally, it is worth stressing that the covariance matrices were computed assuming the fiducial $\Lambda$CDM cosmology. This means that we neglect the error due to the use of a non-$\Lambda$CDM fiducial cosmology. The computation of covariance matrices for alternative cosmologies is indeed an open issue in the exploitation of next-generation survey data \citep[]{Harnois-Deraps:2019rsd,Friedrich:2020dqo}.

\subsection{SNe surveys}

We considered two future SNe surveys, one based on the proposed \Euclid DESIRE survey \citep{Laureijs:2011gra,Astier:2014swa}, and the other based on the specifications of the LSST. We assumed that the \Euclid DESIRE survey will observe $1700$ supernovae in the range $z\in[0.7,1.6]$, while the LSST survey will observe $8800$ supernovae in the redshift range $z\in[0.1,1.0]$ for a total of 10500 data points.

In both cases we assumed the redshift distributions described in \citet{Astier:2014swa}, and we further assumed that they are uncorrelated with each other. While the DESIRE survey is not a guaranteed output of \Euclid, we include it here in order to extend the redshift range of LSST, as this is crucial for the GA reconstruction at high $z$. For every SNe event, we assumed an observational error of the form
\begin{equation}
    \sigma_{\textrm{tot},i}^2=\delta \mu^2_i+\sigma^2_{\textrm{flux}}+\sigma^2_{\textrm{scat}}+\sigma^2_{\textrm{intr}}\,,
\end{equation}
where the flux, scatter, and intrinsic contributions are the same for each event: $\sigma_{\textrm{flux}} = 0.01$, $\sigma_{\textrm{scat}} = 0.025$, and $\sigma_{\textrm{intr}} = 0.12,$ respectively \citep[see][]{Astier:2014swa}. We also included an error on the distance modulus $\mu=m-M_0$ that evolves linearly with the redshift $z$ as
\begin{equation}
    \delta\mu = e_{\rm M}~z\,,
\end{equation}
where we assumed that the parameter $e_{\rm M}$ is drawn from a Gaussian distribution with zero mean and standard deviation $\sigma(e_{\rm M})=0.01$ \citep[see][]{Gong:2009yk,Astier:2014swa}. The distance modulus error takes the possible redshift evolution of SNe into account, which has not been accounted for by the distance estimator; see \citet{Astier:2014swa}. We note that  $e_{\rm M}=0.01$ is needed to allow for systematic evolution, but it would just add in quadrature to the effective $e_{\rm M}=0.055$ coming from lensing, to make a single effective $e_{\rm M}$ that is negligibly different from 0.055.

The uncertainty due to lensing was estimated theoretically to be $\sigma_{\rm lens}= 0.052 z$ \citep[][]{Amendola:2013twa,Quartin:2013moa} and $\sigma_{\rm lens}= 0.056 z$ \citep[][]{Ben-Dayan:2013nkf}, and observationally with Supernova Legacy Survey data to be $\sigma_{\rm lens}= (0.055 \pm 0.04) z$ \citep[][]{Jonsson:2010wx} and $\sigma_{\rm lens}= (0.054 \pm 0.024) z$ \citep[][]{SNLS:2010rmd}.

\subsection{Large-scale structure surveys}

As we are interested in forecasting the sensitivity of the null tests for \lcdm with \Euclid, we now describe how we simulated BAO data using the Fisher matrix approach. To do this, we followed the same method as was used in \citetalias{IST:paper1} for the spectroscopic survey.

We mainly focus on the spectroscopic Euclid survey because our goal is to obtain precise measurements of the Hubble parameter $H(z)$ and the angular diameter distance $D_\mathrm{A}(z)$. Overall, the Euclid survey will be able to probe the galaxy power spectrum in the redshift range $z \in [0.9,1.8]$ where, as mentioned in \citetalias{IST:paper1}, the main targets are ${\rm H}_{\alpha}$ emitters. In this case, \Euclid will measure up to $30$ million spectroscopic redshifts with an uncertainty given by $\sigma_z = 0.001(1 + z)$ \citep{Pozzetti:2016cch} and the main observable will be the galaxy power spectrum. This power spectrum carries information about the distortions due to the Alcock-Paczynski effect, the residual shot noise, the redshift uncertainty, and the anisotropies due to redshift space distortions and on the galaxy bias. Moreover, nonlinear effects that distort the shape of the power spectrum, for instance, a nonlinear smearing
of the BAO feature, were also included in the matter power spectrum \citep{Wang:2012bx}. In principle, other effects might also include a nonlinear scale-dependent galaxy bias;d see for example \citet{delaTorre:2012dg}.

We use the same binning scheme as in \citet{Martinelli:2020hud}, which is different from that of \citetalias{IST:paper1}; specifically, we used nine equally spaced redshift bins of width $\Delta z = 0.1$ instead of four. By rebinning the data as given in \citetalias{IST:paper1}, we find the following specifications for the galaxy number density $n(z)$ in units of Mpc$^{-3}$ and the galaxy bias $b(z)$,
\begin{align}
n(z)&= \{2.04, 2.08, 1.78, 1.58, 1.39, 1.15, 0.97, 0.7, 0.6\}\times 10^{-4}\,,\nonumber \\
b(z)&= \{1.42, 1.5, 1.57, 1.64, 1.71, 1.78, 1.84, 1.90, 1.96\} \,.\nonumber
\end{align}
This binning scheme allows for more data points and considerably improves the machine learning analysis, as discussed below.
However, we note that we tested this particular choice against that of \citetalias{IST:paper1} in \citet{Martinelli:2020hud}, where we found no statistically significant difference.

In order to obtain the Fisher matrix for the full set of cosmological parameters, which is used to estimate the parameter covariance matrix and propagate the error estimates, we followed the procedure described in \citetalias{IST:paper1}. The analysis includes the following parameters: the four shape parameters $\{\omega_{\rm m}=\Omega_{\rm m,0}h^2$, $h$, $\omega_{\rm b}=\Omega_{\rm b,0}h^2$, and $n_{\rm s}\}$, the two nonlinear parameters $\{\sigma_{\rm p}\text{ and}\,\sigma_{\rm v}\}$ (see \citetalias{IST:paper1}), and the five redshift-dependent parameters $\{\ln D_{\rm A},\,\ln H,\,\ln f\sigma_8,\,\ln b\sigma_8,\text{and}\,P_{\rm s}\},$ evaluated in each redshift bin, where $f\sigma_8\equiv f(z)\sigma_8(z)$ is the linear growth rate times $\sigma_8$, which measures the amplitude of the linear power spectrum at scales of $8 h^{-1} \mathrm{Mpc}$, while $b\sigma_8\equiv b(z)\sigma_8(z)$ and $P_{\rm s}$ characterize the galaxy bias and the shot noise, respectively (see \citetalias{IST:paper1}). In this way, we can derive the expected uncertainties from the Euclid survey of the Hubble parameter $H(z)$ and the angular diameter distance $D_{\rm A}(z)$ in each of the nine redshift bins, while we marginalize over all the other parameters. The final Fisher matrix in principle depends on the particular fiducial cosmology, but here we assumed that this dependence is weak at best.

As the \Euclid spectroscopic survey will only probe the redshift range $z \in [0.9,1.8]$, we will be limited in the range in which both BAO and the SNe data are available. Thus, we complemented our analysis by including the DESI survey, in order to be able to reconstruct the null tests at the full range of the available SNe data. DESI has started survey operations in 2021 and is scheduled to obtain optical spectra for tens of millions of quasars and galaxies up to $z\sim 4$, which will allow BAO and redshift-space distortion analyses.

We followed the official DESI forecasts for the Hubble parameter $H(z)$ and the angular diameter distance $D_{\rm A}(z)$ as described in \citet{DESI2016}. These forecasts have also been derived following a Fisher matrix approach, described in \citet{2014JCAP...05..023F}, which is the full anisotropic galaxy power spectrum, that is, measurements of the matter power spectrum as a function of the angle with respect to the line of sight, but also redshift and wavenumber. Similarly to the \Euclid forecasts, this approach also includes all the available information from the two-point correlation function and not just the position of the BAO peak.

\begin{figure*}[!tbp]
\centering
\includegraphics[width = 0.97\textwidth]{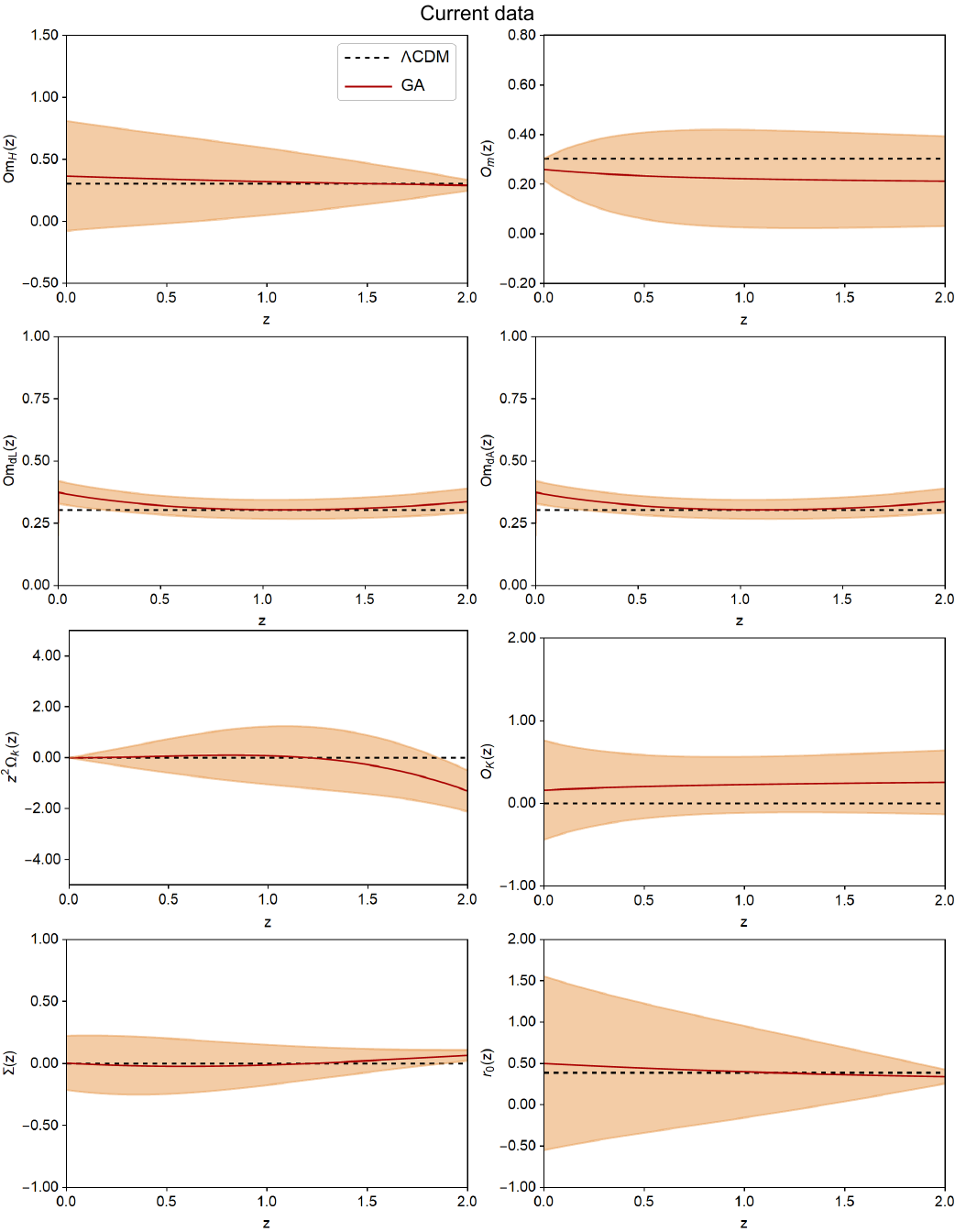}
\caption{GA reconstructions of the null tests mentioned in \Cref{sec:tests} using the currently available BAO and Pantheon SNe data, as described in \Cref{sec:cur_data}. In all cases, the dashed line at zero corresponds to the best-fit $\Lambda$CDM model, described by \Cref{eq:Hlcdm} and the parameters $(\Omega_\mathrm{m,0}=0.303, h=0.659)$, the red line is the GA fit, and the orange shaded region corresponds to the $68\%$ GA uncertainties.}
\label{fig:current}
\vspace*{0cm}
\end{figure*}

Specifically, we considered the baseline DESI survey, covering 14\,000\,deg$^2$ and targeting bright galaxies (BGs), luminous red galaxies (LRGs), emission-line galaxies (ELGs), and quasars in the redshift range $z\in [0.05,3.55]$, but with a precision that will explicitly depend on the target population. First, the BGs will be in the redshift range $z\in [0.05,0.45]$ given in 5 equally spaced redshift bins, while the next two targets, the LRGs and ELGs, will be in the range $z\in [0.65,1.85]$ in $13$ equally spaced redshift bins. Moreover, the Ly-$\alpha$ forest quasars will be given in the range $z\in [1.96,3.55]$ in $11$ equally spaced redshift bins. Finally, we assumed that these measurements are not correlated with each other.

In our analysis we only included the DESI data at late times that do not overlap the \Euclid points in order to avoid spurious correlations between the two surveys. Furthermore, as the SNe data from LSST + DESIRE will only reach redshift $z=1.6$, we only included the DESI data from the full BGs survey and the LRGs and ELGs up to $z=0.9$, thus leaving out the Ly-$\alpha$ forest observations.

\section{Results from the current data\label{sec:currentresults}}

In this section we now present the reconstructions of the null tests of \Cref{sec:tests} using the currently available SNe and BAO data. In this case, as the BAO data are coming from a plethora of different surveys and in a variety of different forms, that is, in terms of $d_z(z)$, $1/d_z(z), \text{and so on,} $  we cannot bin them without introducing new assumptions on their statistical properties. Thus, we only present the results of the GA analysis, following the method presented in \Cref{sec:GA}. In order to simplify the discussion of the results, we split the tests into groups of related tests.

Specifically, we show in \Cref{fig:current} the GA reconstructions of the null tests. In the first row from the top of \Cref{fig:current} we show the GA reconstructions of the two tests for the matter density, which directly probe the Hubble expansion rate, namely $\textrm{Om}_\mathrm{H}(z)$ given by \Cref{eq:omh1} (left) and $\mathcal{O}_{\rm m}(z)$ given by \Cref{Om-hz} (right). They agree with a constant value and the \lcdm within the $68\%$ uncertainties, but $\mathcal{O}_{\rm m}(z)$ is closer to the $68\%$ confidence level boundary because of the presence of the derivatives of the Hubble parameter, which amplify any deviations from the concordance value; see \Cref{Om-hz}.

In the second row from the top, we show the reconstructions for the matter density tests that directly probe the cosmological distances, namely $\textrm{Om}_\textrm{dL}(z)$ given by \Cref{eq:OmdLnull} (left panel) and  $\textrm{Om}_\textrm{dA}(z)$ given by \Cref{eq:OmdAnull} (right panel). They also agree well with a constant value and the \lcdm best fit within the errors. Next, in the third row from the top, we show the two curvature tests that probe the spatial homogeneity of the Universe, namely $z^2 \Omega_{\rm k}(z)$ given by \Cref{eq:OK}  (left panel) versus $\mathcal{O}_{\rm K}(z)$  given by \Cref{OK-hz} (right panel). In this case, both tests are consistent with a flat universe at the $68\%$ confidence level, but the uncertainties seem to increase with redshift for the $z^2 \Omega_{\rm k}(z)$ due to the additional $z^2$ term, which suppresses the singularity at late times, as explained in \Cref{sec:tests}. On the other hand,  $\mathcal{O}_{\rm K}(z)$ is closer to the $68\%$ confidence level boundary because it is complementary to $\mathcal{O}_{\rm m}(z)$ and is subject to the same impact of the presence of the Hubble derivative in their expression.

Furthermore, in the bottom row, we show the global shear test $\Sigma(z)$ (left panel) given by \Cref{eq:GlobalShear}, which tests the Copernican principle, and the $r_0(z)$ null test (right panel), which probes for dark matter-dark energy interactions and is given by \Cref{eq:r0null}. Again, both tests are consistent within the uncertainties with the assumption of the Copernican principle and no interactions in the dark sector. In some cases, for instance, in the $\mathrm{Om}_\mathrm{H}(z)$ and the $r_0(z)$ tests, the error regions increase at low redshifts because these tests are ill defined in the zero redshift limit, as discussed in \Cref{sec:tests}. Similarly, this also occurs for the $z^2\Omega_\mathrm{k}(z)$ test, but because of the additional $z^2$ , which regularizes the singularity at $z=0$, it now instead occurs at high redshifts.

Finally, as the GA does not to evaluate the statistical significance of a potential departure from the \lcdm in a straightforward fashion, we also implemented a quantitative approach of estimating the average deviation from the null hypothesis and the average size of the errors across the redshift range of the data. We also present the results of the mock data below, and then this approach is particularly useful because it allows us to obtain an overall improvement factor that forthcoming surveys will add when they are compared to the current data of this section. The technical details of this analysis are given in \Cref{app:quantimp}.

\section{Forecast results\label{sec:fore_res}}
Following the approach described in \Cref{sec:fore_data}, we now present the constraint estimates of our null tests using the SNe and BAO mock data for three different fiducial cosmologies based on the vanilla flat \lcdm model given by \Cref{eq:Hlcdm}, the CPL parameterization given by \Cref{eq:CPLpar}, and the \lltb model described by \Cref{eq:LLTB_fried}, assuming the fiducial values for the parameters shown in \Cref{tab:fiducials}. In our analysis we also consider two different cases. In the first case, we only use the \Euclid BAO data, spanning the range $z\in[0.9,1,8]$, in order to quantify the ability of the Euclid survey alone to constrain any deviations from the null tests. In a second case, we also include the DESI BAO data, which cover the lower redshifts, as discussed in \Cref{sec:fore_data}, in order to highlight the synergies of the two surveys. Both cases include all SNe data (\Euclid + LSST), and similarly to \Cref{sec:currentresults}, we group the plots by row, splitting the tests into groups of related tests.

As in this case the mock BAO data are always given in terms of the angular diameter distance $D_\mathrm{A}(z)$ and the Hubble expansion rate $H(z)$, we are now also able to perform the binning analysis as discussed in \Cref{sec:parapp} as a model-independent analysis complementary to that of the GA. The binned data produced by this approach are also used to obtain constraint estimates on the considered tests, using a parametric approach based on a CPL-like parameterization given by \Cref{eq:CPL}. However, that while we apply the GA simultaneously to the full SNe and BAO mock data, the PA is only applied on the binned data of each test.

Finally, as mentioned in the previous section, we also implement a quantitative approach of estimating the overall improvement factors that are expected from forthcoming surveys when compared to the current data of the previous section. The technical details of this analysis are given in \Cref{app:quantimp}.

\subsection{Results from the \lcdm mocks}
\begin{figure*}[!htbp]
\centering
\includegraphics[width = 0.975\textwidth]{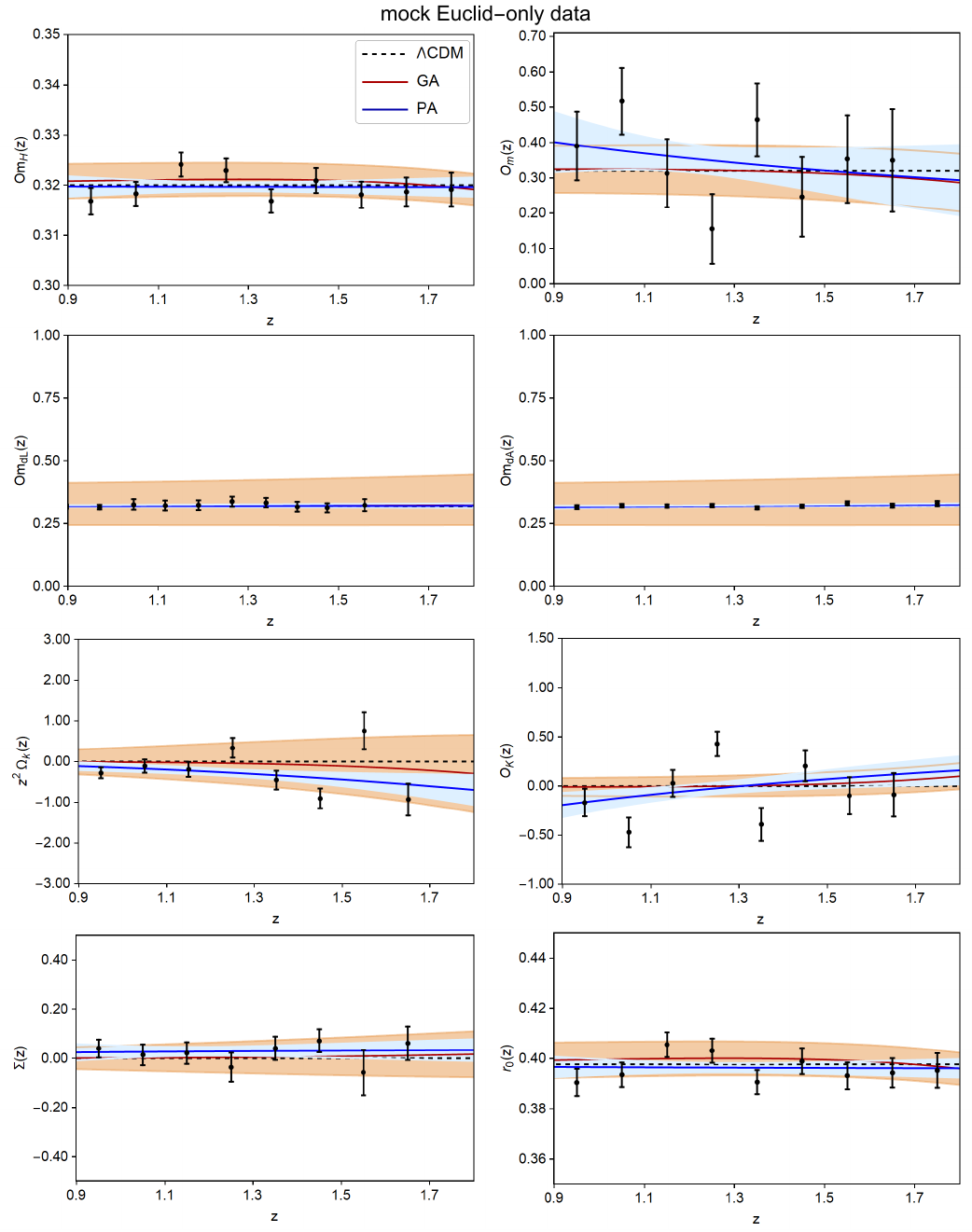}
\caption{GA reconstructions of the null tests mentioned in \Cref{sec:tests} using the \Euclid-only BAO and \Euclid + LSST SNe \lcdm mock data. In all cases, the dashed line corresponds to the fiducial value of the corresponding test for $\Lambda$CDM, the red line is the GA fit, the blue line is the PA fit, the shaded regions correspond to the $68\%$ uncertainties, and the black points correspond to the binned \Euclid mock data.}
\label{fig:mockGAEuclid}
\end{figure*}

\begin{figure*}[!htbp]
\centering
\includegraphics[width = 0.975\textwidth]{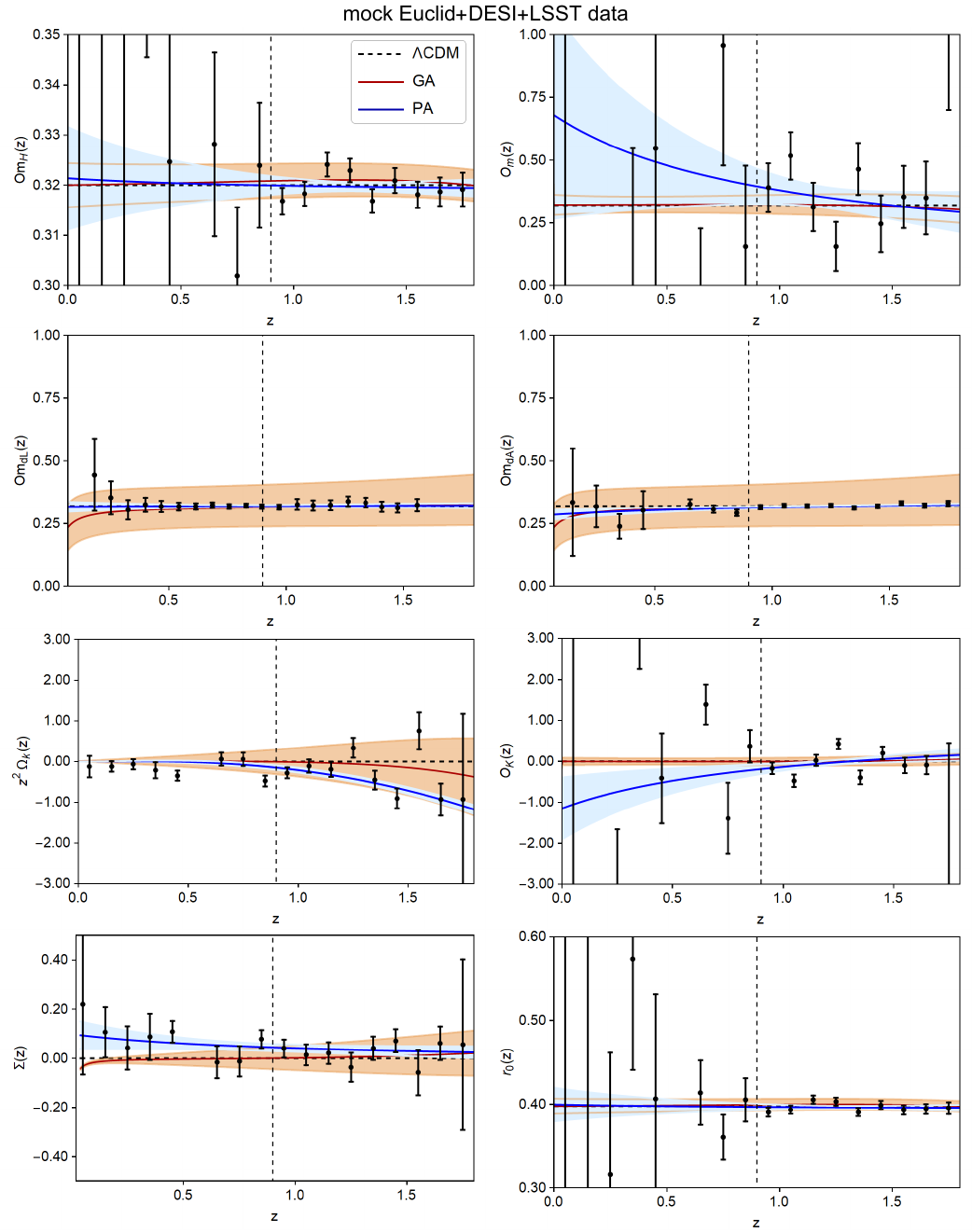}
\caption{GA reconstructions of the null tests mentioned in \Cref{sec:tests} using the \Euclid + DESI BAO and \Euclid + LSST SNe \lcdm mock data. In all cases, the dashed line corresponds to the fiducial value of the corresponding test for $\Lambda$CDM, the red line is the GA fit, the blue line is the PA fit, the shaded regions correspond to the $68\%$ uncertainties, and the black points correspond to the binned \Euclid and DESI mock data. The vertical dashed line at $z=0.9$ indicates the minimum redshift of the \Euclid-only points.}
\label{fig:mockGAall}
\end{figure*}

First, we present in \Cref{fig:mockGAEuclid} the results of the GA reconstructions of the null tests described in \Cref{sec:tests} for the \Euclid-only \lcdm mocks and in \Cref{fig:mockGAall} for all mock data (\Euclid + DESI). In all cases, both here and in the plots below, the dashed line corresponds to the fiducial value of the test under consideration for $\Lambda$CDM, the red line is the GA fit, the blue line is the PA fit given by \Cref{eq:CPL} of the binned mock data (black points), while the orange and blue shaded regions correspond to the $68\%$ uncertainties of the GA and PA, respectively, and the vertical dashed line at $z=0.9$ indicates the minimum redshift of the \Euclid-only points. Because the binned mock data include a random realization of error, they are expected to fluctuate around the fiducial value. This is indeed seen in \Cref{fig:mockGAEuclid} and \Cref{fig:mockGAall}.

All of the panels of \Cref{fig:mockGAEuclid} show that in all cases, both the GA and the PA reconstructions along with the binned mock data are able to correctly predict the fiducial model to within the $95\%$ uncertainties. Because of their agnostic nature, the uncertainties of the GA reconstructions (orange shaded regions) are somewhat larger than those of the PA (blue shaded regions), but they are very similar to those of the binning (black points). An exception to this is the $\text{Om}_\mathrm{dA}(z)$ test, which in the case of the GA is dominated by the systematic uncertainties of the SNe data because of the joint reconstruction approach employed in this case, as was also observed in \citet[]{Martinelli:2020hud}. In particular, as we fit all the data (SNe and BAO) simultaneously, the errors of the GA reconstructions for the distances are dominated by the measurements with the larger (worst) errors, which in this case are the SNe. Hence, using the GA reconstructions for the angular diameter distance results in much larger errors than in the PA method. The PA method does not suffer from this issue as the PA fits the Taylor expansion directly only to the binned BAO angular diameter data.

On the other hand, when we also include the DESI BAO data at low and intermediate redshifts that do not overlap the \Euclid points, as shown in \Cref{fig:mockGAall}, we find that the uncertainties in the case of the binning approach dramatically increase because the constraining power of DESI is lower than that of \Euclid at this redshift range. This affects the PA (blue shaded regions), as in some cases, such as for the $\Sigma(z)$ and the $\mathcal{O}_{\rm K}(z)$ tests, it misses the fiducial model (dashed line) by more than $1\sigma$. The reason for this is that the PA is anchored to the \Euclid points at intermediate redshifts, which have smaller error bars than the \Euclid redshifts, and because the PA parameterization lacks flexibility, discrepancies appear at low redshifts. This also affects the GA reconstruction for $\Sigma(z)$, but in the remaining cases, the GA efficiently predicts the correct cosmology as it uses the full SNe and BAO data. In either case, these deviations are due to the presence of derivatives in the null tests, which tend to cause instabilities in the reconstructions.

Overall, after analyzing the \lcdm mocks, we find that \Euclid, in combination with other surveys, will be able to improve current constraints by approximately a factor of three with the machine learning approach. The binning and parametric approach will provide an improvement of a further factor of two over the GA results.

\subsection{Results from the CPL mocks}
\begin{figure*}[!htbp]
\centering
\includegraphics[width = 0.975\textwidth]{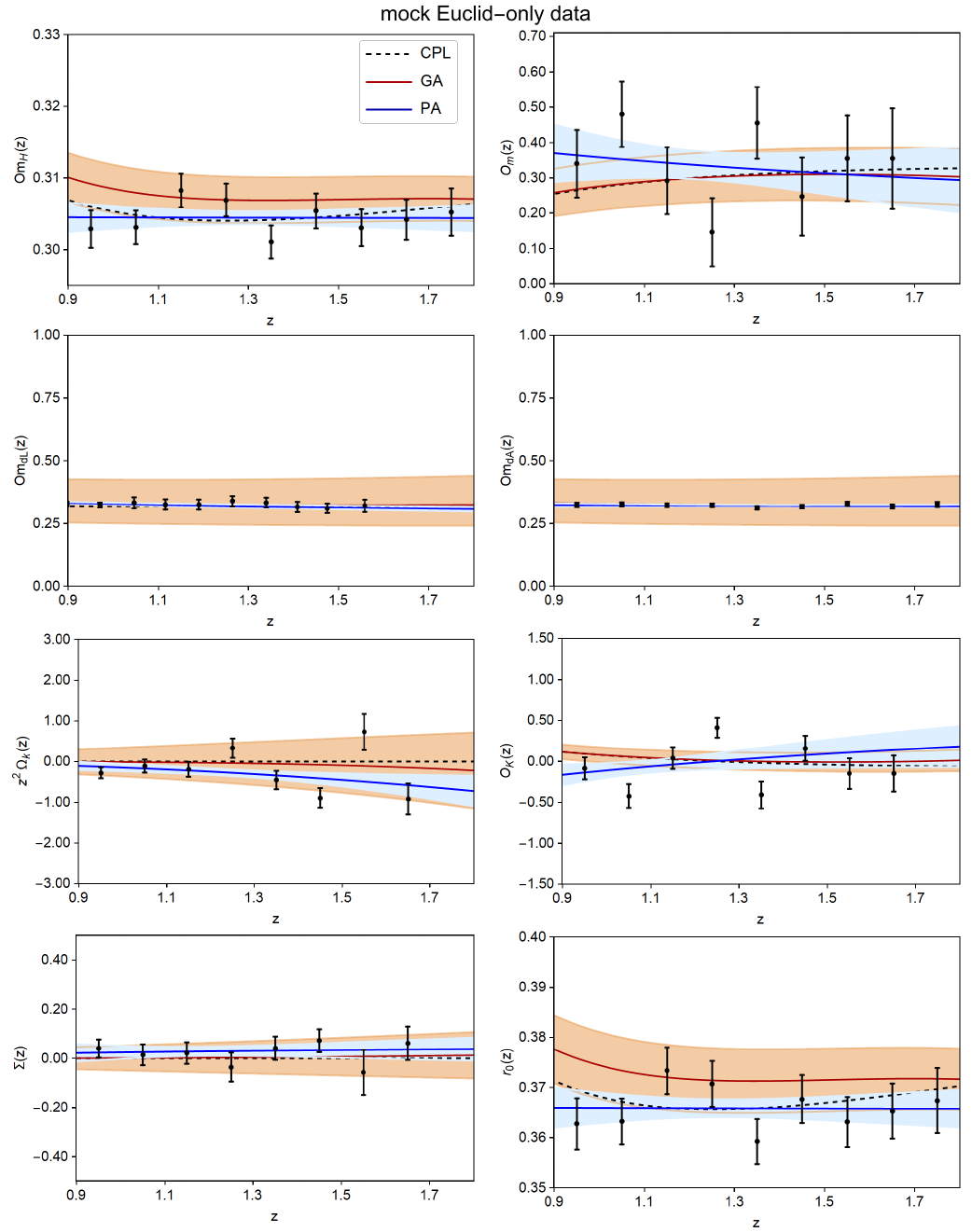}
\caption{GA reconstructions of the null tests mentioned in \Cref{sec:tests} using the \Euclid-only BAO and \Euclid + LSST SNe data for the CPL mock for $(w_0,w_a)=(-0.8,-1)$. In all cases, the dashed line corresponds to the fiducial value of the corresponding test for CPL, the red line is the GA fit, the blue line is the PA fit, the shaded regions correspond to the $68\%$ uncertainties, and the black points correspond to the binned \Euclid  mock data.}
\label{fig:mockCPLGAEuclid}
\end{figure*}

\begin{figure*}[!htbp]
\centering
\includegraphics[width = 0.975\textwidth]{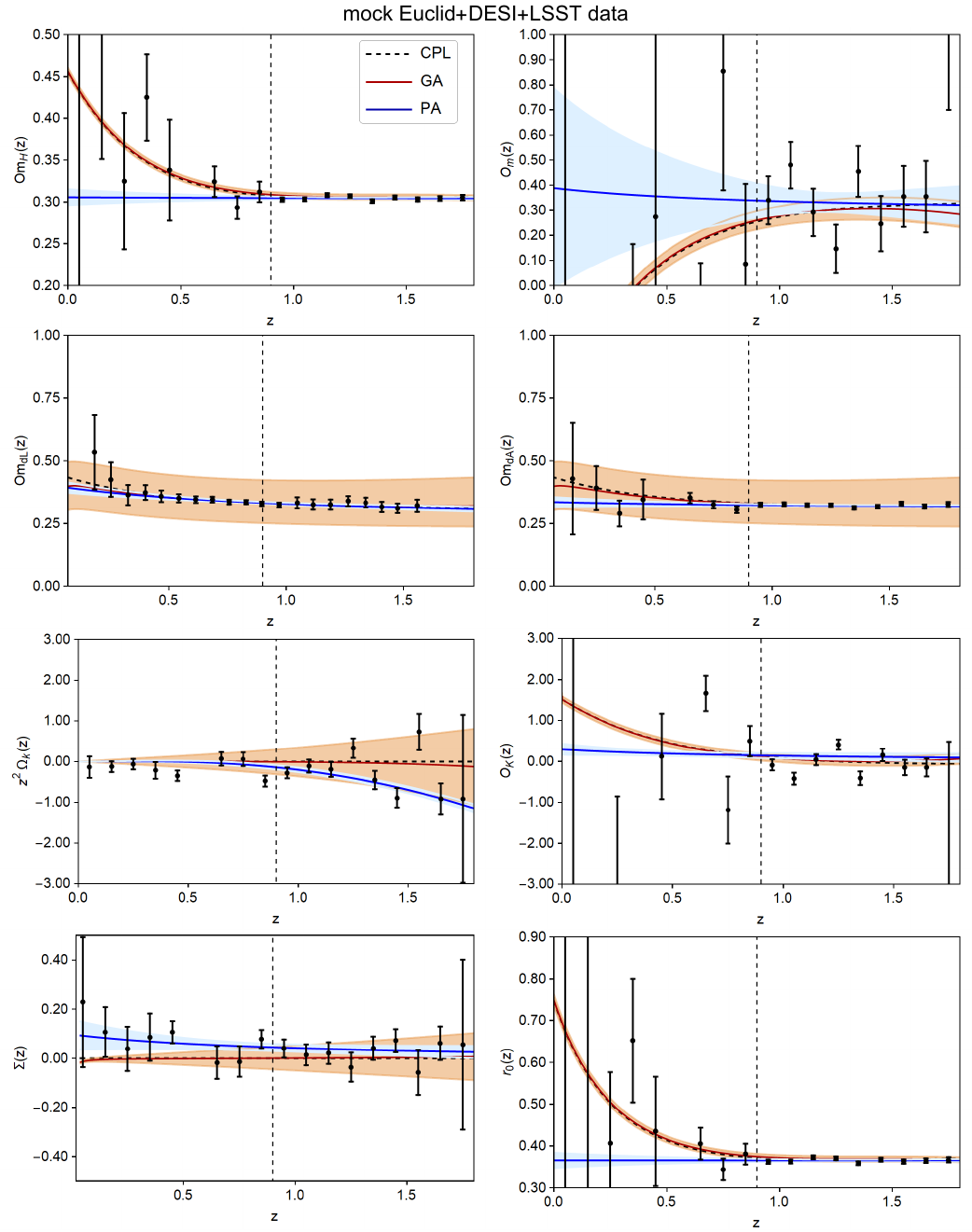}
\caption{GA reconstructions of the null tests mentioned in \Cref{sec:tests} using the \Euclid + DESI BAO and \Euclid + LSST SNe data for the CPL mock for $(w_0,w_a)=(-0.8,-1)$. In all cases, the dashed line corresponds to the fiducial value of the corresponding test for CPL, the red line is the GA fit, the blue line is the PA fit, the shaded regions correspond to the $68\%$ uncertainties, and the black points correspond to the binned \Euclid and DESI mock data. The vertical dashed line at $z=0.9$ indicates the minimum redshift of the \Euclid-only points.}
\label{fig:mockCPLGAall}
\end{figure*}

Next, we present in \Cref{fig:mockCPLGAEuclid} the results of the GA and parametric reconstructions of the null tests for the \Euclid-only CPL mocks. In \Cref{fig:mockCPLGAall} we show them for all mock data (\Euclid + DESI).

This case now is more interesting because the nature of the DE model we chose means that many of the tests that probe the expansion history of the Universe, such as the $\mathrm{Om}_\mathrm{H}(z)$, have a distinct evolution with redshift. In the case of the \Euclid-only mocks shown in \Cref{fig:mockCPLGAEuclid},  both approaches overall predict the fiducial cosmology within the errors, but in some cases, there is a small $1\sigma$ deviation. This occurs, for example, in the case of the $\mathrm{Om}_\mathrm{H}(z)$  test with the GA because the GA cannot anchor very well at $z\sim 1$ because it lacks points at low redshift. On the other hand, in the case of the binning, for the $\mathcal{O}_{\rm m}(z)$ and  $\mathcal{O}_{\rm K}(z)$ tests, the presence of derivatives in the expressions causes more scatter in the points, which is also propagated to the PA (blue line). This also affects the $r_0(z)$ test, where the effect is further amplified by the need to fix the baryon density parameter $\Omega_\textrm{b,0}$ to its {\it Planck} best-fit value.

When we also include the DESI BAO data, see \Cref{fig:mockCPLGAall}, we can extend the reconstructions of the tests to low redshifts, where many of the tests, such as the fiducial value of the $\mathrm{Om}_\mathrm{H}(z)$ test, manifest large deviations from a constant value. In this case, we find that while the GA follows the fiducial trend quite well, the PA (blue line) is anchored to the higher redshift \Euclid points, as these have smaller uncertainties than the DESI points, and it deviates from the fiducial model by several $\sigma$; a similar effect is also visible for the $r_0(z)$ test. Finally, neither in \Euclid-only nor in \Euclid + DESI BAO data is the global shear test affected as expected, and the reconstructions are practically identical to the reconstruction of the \lcdm mocks.

\begin{figure*}[!htbp]
\centering
\includegraphics[width = 0.975\textwidth]{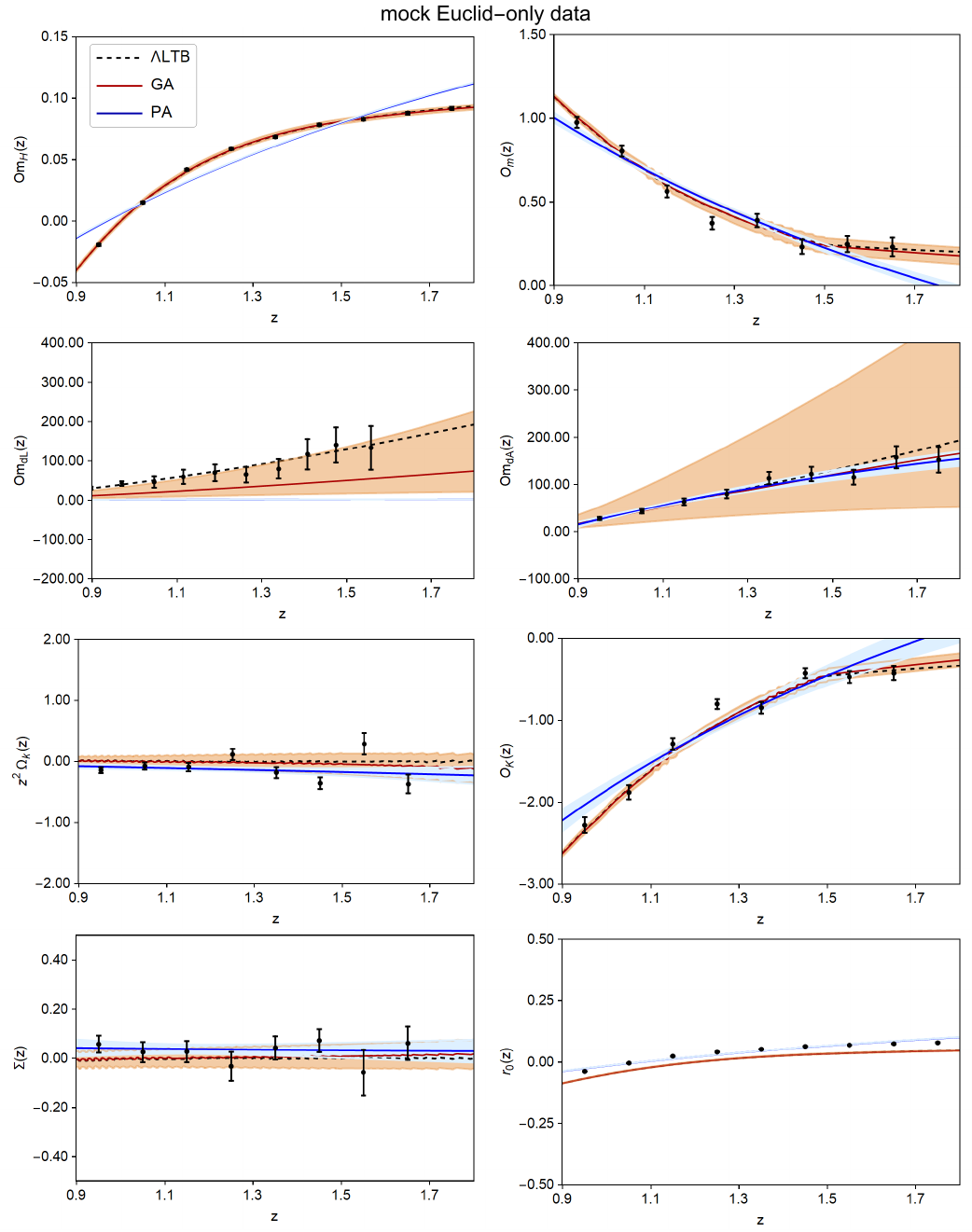}
\caption{GA reconstructions of the null tests mentioned in \Cref{sec:tests} using the \Euclid-only BAO and \Euclid + LSST SNe data for the \lltb mock. In all cases, the dashed line corresponds to the fiducial value of the corresponding test for the \lltb model, the red line is the GA fit, the blue line is the PA fit, the shaded regions correspond to the $68\%$ uncertainties, and the black points correspond to the binned \Euclid mock data.}
\label{fig:mockLTBGAEuclid}
\end{figure*}

\begin{figure*}[!htbp]
\centering
\includegraphics[width = 0.975\textwidth]{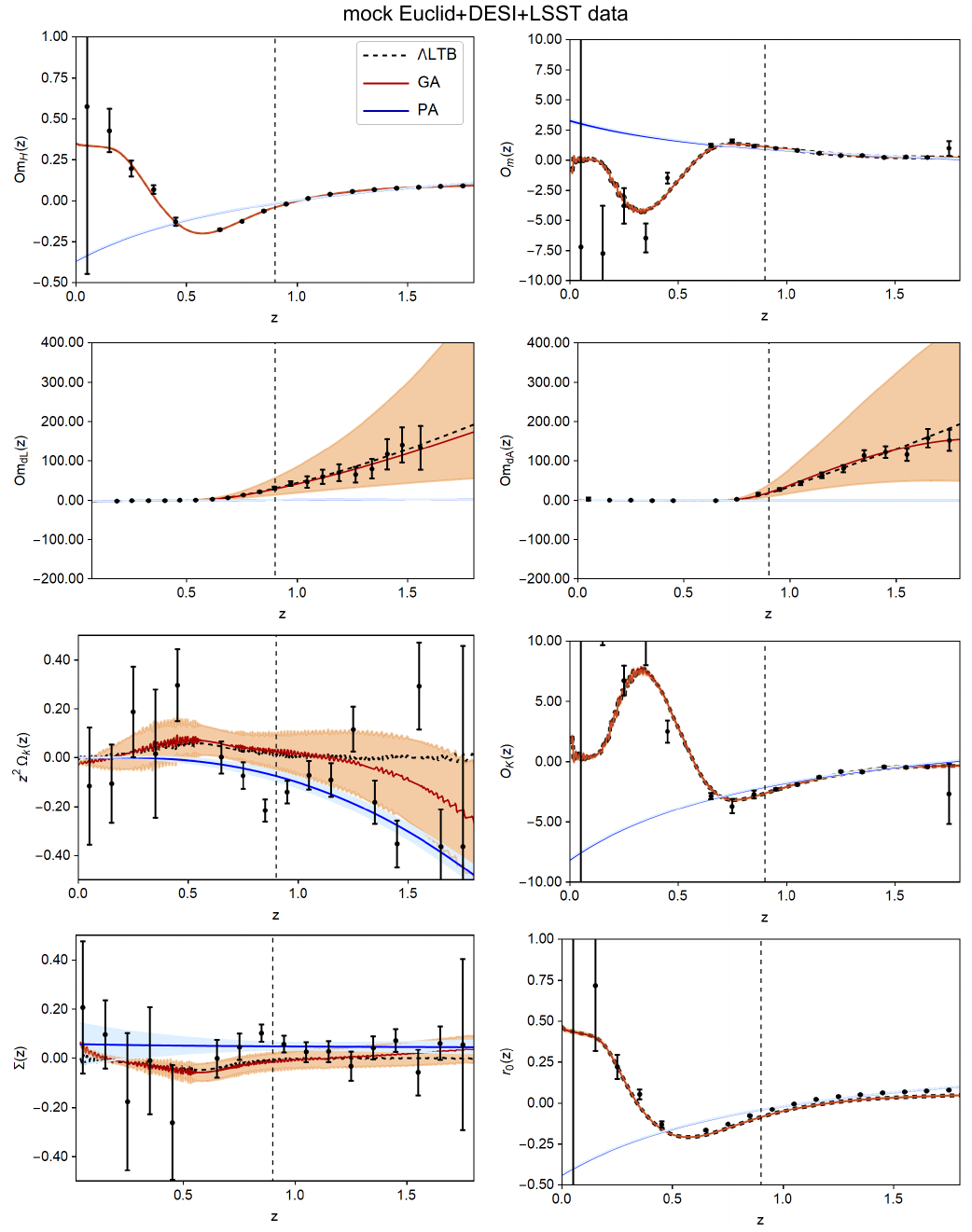}
\caption{GA reconstructions of the null tests mentioned in \Cref{sec:tests} using the \Euclid + DESI BAO and \Euclid + LSST SNe data for the \lltb mock. In all cases, the dashed line corresponds to the fiducial value of the corresponding test for the \lltb model, the red line is the GA fit, the blue line is the PA fit, the shaded regions correspond to the $68\%$ uncertainties, and the black points correspond to the binned \Euclid and DESI mock data. The vertical dashed line at $z=0.9$ indicates the minimum redshift of the \Euclid-only points.}
\label{fig:mockLTBGAall}
\end{figure*}

\subsection{Results from the \lltb mocks}
Finally, we now present in \Cref{fig:mockLTBGAEuclid} the results of the GA and PA for the \Euclid-only \lltb mocks and in \Cref{fig:mockLTBGAall} for all mock data (\Euclid + DESI). This particular case is even more interesting, as all the tests present large deviations compared to the simple \lcdm case. This is due to the void profile we used. Specifically, when compared to the corresponding plots for the \lcdm mocks, the \lltb mocks exhibit large deviations from a constant value at small redshifts, due to the shape of the void profile, in particular in the cases of $\mathrm{Om}_\mathrm{H}(z)$, $r_0(z)$, $\mathcal{O}_\mathrm{m}(z)$ and $\mathcal{O}_\mathrm{K}(z)$.

\Cref{fig:mockLTBGAEuclid} shows that in the case of the \Euclid-only BAO data, the PA given by \Cref{eq:CPL} is not flexible enough to reproduce the fiducial model; see, for example, the top left panel of \Cref{fig:mockLTBGAEuclid} for the $\mathrm{Om}_\mathrm{H}(z)$ reconstruction or the $\mathrm{Om}_\mathrm{dL}(z)$ and $\mathrm{Om}_\mathrm{dA}(z)$ tests in the second row. The latter actually give very high values of the matter density. This would be a smoking gun for the presence of voids. In the case of the $\mathcal{O}_{\rm K}(z)$ test, the GA is much closer to the fiducial model than to the PA, while for the $r_0(z)$ test, the deviation is again partly due to fixing the value of the baryon density, as before. When we also include the DESI BAO data, the reconstructions of \Cref{fig:mockLTBGAall} clearly highlight the ability of our approach to fully harness the power of the null tests of \Cref{sec:tests}.

In particular, because of the choice of the \lltb profile, several of the tests, namely $\mathrm{Om}_\mathrm{H}(z)$, $\mathcal{O}_{\rm K}(z)$, $r_0(z)$, and $\Sigma(z)$, have large deviations from a constant value, which the GA is able to capture correctly. The simultaneous fitting of the data with the GA, see \Cref{sec:GA}, causes some correlations between the Hubble parameter and the luminosity distance, however. As a result, the fits in the redshift range covered by \Euclid might differ between the \Euclid-only and \Euclid + DESI cases because of the added DESI BAO data at low $z$ in the latter case, which may shift the fit. This difference in the fits between the two data sets can be seen, for example, in the $\mathrm{Om}_\mathrm{dL}(z)$ test shown in \Cref{fig:mockLTBGAEuclid} and  \Cref{fig:mockLTBGAall} for the two sets of mocks.

On the other hand, in these cases, the PA is again anchored at high redshifts because the uncertainties of the \Euclid data are small, and because of its simple redshift evolution, it misses the features at small $z$. The same but in the opposite direction occurs for the reconstructions of the $\mathrm{Om}_\mathrm{dL}(z)$ and $\mathrm{Om}_\mathrm{dA}(z)$ tests, but in this case, the GA and the parametric approach are always within the errors.

\section{Conclusions \label{sec:conclusions}}
We have constrained deviations from the Copernican principle, the spatial homogeneity of the Universe on large scales, and from the \lcdm model using a set of null tests, that is, identities that have to be equal to a constant value at all redshifts, using current and forecast data. For the latter, we mainly focused on the constraints coming from synergies between \Euclid and other surveys, namely DESI and LSST, using BAO and SNe data.

In \Cref{sec:theory} we presented several mechanisms for which fundamental assumptions of the standard cosmological model, such as spatial homogeneity and the Copernican principle, may break down, as in the case of Bianchi or LTB models, for instance. For this reason, we employed two theory-agnostic reconstruction approaches, one approach based on machine learning, called the GA, and another based on binning the data and then fitting a simple CPL-like parameterization, which we called the PA. These approaches allowed us to derive constraints on any deviation from the standard model with only a few assumptions on the redshift trends of these deviations. Avoiding any further assumptions is particularly important as assuming a particular model and then fitting the data introduces theoretical biases, and as we have shown, makes it prone to misinterpreting deviations from the standard model.

In order to probe for deviations from the Copernican principle, the spatial homogeneity of the Universe on large scales, and from the \lcdm model we used a set of null tests proposed over the years in the literature, which we summarized in \Cref{sec:tests}. These tests probe different aspects of the aforementioned assumptions and provide a holistic approach to quantifying whether we might be able to detect any deviations from them.

We first considered the currently available SNe and BAO data. Using the GA, we found that within the errors, all the tests are compatible with the null hypothesis, that is, there were no deviations from the assumptions of the standard cosmological model, see \Cref{sec:currentresults} and \Cref{fig:current}. This is to be expected to some extent because the uncertainties of the currently available SNe and BAO data are large, which makes it difficult to detect any trends in the underlying cosmological model.

The fact that our current constraints are limited by the available BAO data highlights the importance of future full-sky surveys, such as \Euclid. Thus, we  considered mock data based on the specifications of \Euclid and also allowed for synergies with DESI and LSST, in order to forecast deviations from the assumptions of the standard cosmological model for three different fiducial cosmologies: one cosmology based on the \lcdm model, one based on the CPL model for parameters at the $95\%$ limit of the current Planck best-fit, and finally, another mock based on the \lltb model, which describes an inhomogeneous universe with a cosmological constant, as described in \Cref{sec:fore_data}.

Performing a simultaneous fit of these mock SNe and BAO data with the GA, we found, using the quantitative method described in \Cref{app:quantimp}, an improvement of a factor of three in each case over the current constraints for most of the consistency tests, assuming the \lcdm mocks, as seen in \Cref{tab:factors}. Alongside the GA, we also applied the PA, which relies on the choice of a parameterization for the trend in redshift of the tests. Overall, we find that the GA provides larger errors than the PA, and this is for two main reasons. First, the GA is a nonparametric approach, thus is more agnostic, and it spans or explores the much larger functional space, instead of just the usual parameter space as in MCMCs. This tends to increase the errors; see \citet[]{Nesseris:2012tt}.

Second, the choice of a parameterization for the PA creates a lack of flexibility, which is particularly evident in the cases of the CPL and \lltb mocks. This results in reconstructions that deviate significantly from expectations. This lack of flexibility also results in error bars that are tighter than those of the GA because the posterior of PA parameters drops sharply moving away from their best fit. These issues can in principle be avoided by increasing the flexibility of the PA, for example, considering higher-order redshift terms in \Cref{eq:CPL}. However, in realistic situations, the degree of complexity needed for the reconstruction would be unknown, and we decided here to use a common way to parameterize a redshift trend. This stresses even more strongly the advantage of the GA, which does not require any a priori assumption to obtain the required level of flexibility.

Finally, we limited our analysis to the $\Lambda$CDM, CPL, and \lltb because it is currently not feasible to go beyond these models, for example, by considering extensions such as the Bianchi type-I or back-reaction models. We lack validated codes that might be used to estimate the necessary functions (distances and expansion rates). It is therefore quite difficult to test the response of our null tests to these extensions of the standard cosmological model, but with new developments in theoretical model simulation, we could explore them in the future.

In summary, this work highlights the benefits of synergies between the \Euclid BAO survey and external probes in constraining  any deviations from the fundamental assumptions of the standard cosmological model. In particular, we have demonstrated that such a BAO survey will make it possible to constrain deviations from the Copernican principle, the spatial homogeneity of the Universe, and the \lcdm model at an unprecedented level in the near future using both non-parametric and parameterized approaches.

\begin{acknowledgements}
We are grateful to S.~\'Avila for interesting discussions. JGB, MM and SN acknowledge support from the research project  PGC2018-094773-B-C32, and the Centro de Excelencia Severo Ochoa Program SEV-2016-059. SN also acknowledges support from the Ram\'{o}n y Cajal program through Grant No. RYC-2014-15843.  MM also received  support from ``la Caixa'' Foundation (ID 100010434), with fellowship code LCF/BQ/PI19/11690015. DS acknowledges financial support from the Fondecyt Regular project number 1200171. The work of CJM was financed by FEDER -- Fundo Europeu de Desenvolvimento Regional funds through the COMPETE 2020 -- Operational Programme for Competitiveness and Internationalisation (POCI), and by Portuguese funds through FCT - Funda\c c\~ao para a Ci\^encia e a Tecnologia in the framework of the project POCI-01-0145-FEDER-028987.
DC thanks CAPES for financial support.
VM thanks CNPq and FAPES for partial financial support.
LL was supported by a Swiss National Science Foundation (SNSF) Professorship grant (No.~170547).
This project has received funding from the European Union’s Horizon 2020 research and innovation programme under the Marie Sklodowska-Curie grant agreement No 888258.
CC is supported by the UK Science \& Technology Facilities Council (STFC) Consolidated Grant ST/P000592/1.
%
AdS acknowledges support from the Fundação para a Ciência e a Tecnologia (FCT) through the Investigador FCT Contract No. IF/01135/2015 and POCH/FSE (EC) and in the form of an exploratory project with the same reference.
JPM and AdS acknowledge support from FCT Projects with references EXPL/FIS-AST/1368/2021, PTDC/FIS-AST/0054/2021, UIDB/04434/2020, UIDP/04434/2020, CERN/FIS-PAR/0037/2019, PTDC/FIS-OUT/29048/2017.
ZS acknowledges support from the IRAP and IN2P3 Lyon computing centers.
PF received the support of a fellowship from ``la Caixa'' Foundation (ID 100010434). The fellowship code is LCF/BQ/PI19/11690018.
IT acknowledges support from the Spanish Ministry of Science, Innovation and Universities through grant ESP2017-89838, and the H2020 programme of the European Commission through grant 776247.

\AckEC
\end{acknowledgements}

\bibliographystyle{aa}
\bibliography{references} 

\providecommand*\hyphen{-}
\begin{thebibliography}{110}
\expandafter\ifx\csname natexlab\endcsname\relax\def\natexlab#1{#1}\fi

\bibitem[{Abbott {et~al.}(2021)}]{DES:2021esc}
Abbott, T. M.~C. {et~al.} 2021 [\eprint[arXiv]{2107.04646}]

\bibitem[{Aizpuru {et~al.}(2021)Aizpuru, Arjona, \& Nesseris}]{Aizpuru:2021vhd}
Aizpuru, A., Arjona, R., \& Nesseris, S. 2021, Phys. Rev. D, 104, 043521

\bibitem[{Akrami {et~al.}(2010)Akrami, Scott, Edsjo, Conrad, \&
  Bergstrom}]{Akrami:2009hp}
Akrami, Y., Scott, P., Edsjo, J., Conrad, J., \& Bergstrom, L. 2010, JHEP, 04,
  057

\bibitem[{Alam {et~al.}(2021)}]{eBOSS:2020yzd}
Alam, S. {et~al.} 2021, Phys. Rev. D, 103, 083533

\bibitem[{{Albrecht} {et~al.}(2006){Albrecht}, {Bernstein}, {Cahn}, {Freedman},
  {Hewitt}, {Hu}, {Huth}, {Kamionkowski}, {Kolb}, {Knox}, {Mather}, {Staggs},
  \& {Suntzeff}}]{DEtaskforce}
{Albrecht}, A., {Bernstein}, G., {Cahn}, R., {et~al.} 2006, arXiv e-prints,
  arXiv:astro\hyphen ph/0609591

\bibitem[{Alnes {et~al.}(2006)Alnes, Amarzguioui, \& Gron}]{Alnes:2005rw}
Alnes, H., Amarzguioui, M., \& Gron, O. 2006, Phys. Rev. D, 73, 083519

\bibitem[{Alonso {et~al.}(2015)Alonso, Salvador, S\'anchez, Bilicki,
  Garc\'\i{}a-Bellido, \& S\'anchez}]{Alonso:2014xca}
Alonso, D., Salvador, A.~I., S\'anchez, F.~J., {et~al.} 2015, Mon. Not. Roy.
  Astron. Soc., 449, 670

\bibitem[{{Amendola} {et~al.}(2018){Amendola}, {Appleby}, {Avgoustidis},
  {Bacon}, {Baker}, {Baldi}, {Bartolo}, {Blanchard}, {Bonvin}, {Borgani},
  {Branchini}, {Burrage}, {Camera}, {Carbone}, {Casarini}, {Cropper}, {de
  Rham}, {Dietrich}, {Di Porto}, {Durrer}, {Ealet}, {Ferreira}, {Finelli},
  {Garc{\'\i}a-Bellido}, {Giannantonio}, {Guzzo}, {Heavens}, {Heisenberg},
  {Heymans}, {Hoekstra}, {Hollenstein}, {Holmes}, {Hwang}, {Jahnke},
  {Kitching}, {Koivisto}, {Kunz}, {La Vacca}, {Linder}, {March}, {Marra},
  {Martins}, {Majerotto}, {Markovic}, {Marsh}, {Marulli}, {Massey}, {Mellier},
  {Montanari}, {Mota}, {Nunes}, {Percival}, {Pettorino}, {Porciani},
  {Quercellini}, {Read}, {Rinaldi}, {Sapone}, {Sawicki}, {Scaramella},
  {Skordis}, {Simpson}, {Taylor}, {Thomas}, {Trotta}, {Verde}, {Vernizzi},
  {Vollmer}, {Wang}, {Weller}, \& {Zlosnik}}]{ReviewDoc}
{Amendola}, L., {Appleby}, S., {Avgoustidis}, A., {et~al.} 2018, Living Rev.
  Rel., 21, 2

\bibitem[{Anderson {et~al.}(2014)}]{BOSS:2013rlg}
Anderson, L. {et~al.} 2014, Mon. Not. Roy. Astron. Soc., 441, 24

\bibitem[{Andersson \& Coley(2011)}]{Andersson:2011za}
Andersson, L. \& Coley, A. 2011, Class. Quant. Grav., 28, 160301

\bibitem[{Angulo {et~al.}(2008)Angulo, Baugh, Frenk, \& Lacey}]{Angulo:2007fw}
Angulo, R., Baugh, C., Frenk, C., \& Lacey, C. 2008, Mon. Not. Roy. Astron.
  Soc., 383, 755

\bibitem[{Anselmi {et~al.}(2018)Anselmi, Starkman, Corasaniti, Sheth, \&
  Zehavi}]{Anselmi:2017cuq}
Anselmi, S., Starkman, G.~D., Corasaniti, P.-S., Sheth, R.~K., \& Zehavi, I.
  2018, Phys. Rev. Lett., 121, 021302

\bibitem[{Arjona(2020)}]{Arjona:2020doi}
Arjona, R. 2020, JCAP, 08, 009

\bibitem[{Arjona \& Nesseris(2020{\natexlab{a}})}]{Arjona:2020kco}
Arjona, R. \& Nesseris, S. 2020{\natexlab{a}}, JCAP, 11, 042

\bibitem[{Arjona \& Nesseris(2020{\natexlab{b}})}]{Arjona:2019fwb}
Arjona, R. \& Nesseris, S. 2020{\natexlab{b}}, Phys. Rev. D, 101, 123525

\bibitem[{Arjona \& Nesseris(2021)}]{Arjona:2021zac}
Arjona, R. \& Nesseris, S. 2021, Phys. Rev. D, 104, 103532

\bibitem[{{Astier} {et~al.}(2014){Astier}, {Balland}, {Brescia}, {Cappellaro},
  {Carlberg}, {Cavuoti}, {Della Valle}, {Gangler}, {Goobar}, {Guy}, {Hardin},
  {Hook}, {Kessler}, {Kim}, {Linder}, {Longo}, {Maguire}, {Mannucci},
  {Mattila}, {Nichol}, {Pain}, {Regnault}, {Spiro}, {Sullivan}, {Tao},
  {Turatto}, {Wang}, \& {Wood-Vasey}}]{Astier:2014swa}
{Astier}, P., {Balland}, C., {Brescia}, M., {et~al.} 2014, \aap, 572, A80

\bibitem[{Aubourg {et~al.}(2015)}]{Aubourg:2014yra}
Aubourg, E. {et~al.} 2015, Phys. Rev. D, 92, 123516

\bibitem[{Ben-Dayan {et~al.}(2013)Ben-Dayan, Gasperini, Marozzi, Nugier, \&
  Veneziano}]{Ben-Dayan:2013nkf}
Ben-Dayan, I., Gasperini, M., Marozzi, G., Nugier, F., \& Veneziano, G. 2013,
  JCAP, 06, 002

\bibitem[{Beutler {et~al.}(2011)Beutler, Blake, Colless, Jones, Staveley-Smith,
  Campbell, Parker, Saunders, \& Watson}]{Beutler:2011hx}
Beutler, F., Blake, C., Colless, M., {et~al.} 2011, \mnras, 416, 3017

\bibitem[{Biswas \& Notari(2008)}]{Biswas:2007gi}
Biswas, T. \& Notari, A. 2008, JCAP, 06, 021

\bibitem[{Biswas {et~al.}(2010)Biswas, Notari, \& Valkenburg}]{Biswas:2010xm}
Biswas, T., Notari, A., \& Valkenburg, W. 2010, JCAP, 11, 030

\bibitem[{{Blake} {et~al.}(2012){Blake}, {Brough}, {Colless}, {Contreras},
  {Couch}, {Croom}, {Croton}, {Davis}, {Drinkwater}, {Forster}, {Gilbank},
  {Gladders}, {Glazebrook}, {Jelliffe}, {Jurek}, {Li}, {Madore}, {Martin},
  {Pimbblet}, {Poole}, {Pracy}, {Sharp}, {Wisnioski}, {Woods}, {Wyder}, \&
  {Yee}}]{Blake:2012pj}
{Blake}, C., {Brough}, S., {Colless}, M., {et~al.} 2012, \mnras, 425, 405

\bibitem[{Blanchard {et~al.}(2020)}]{IST:paper1}
Blanchard, A. {et~al.} 2020, Astron. Astrophys., 642, A191

\bibitem[{Bogdanos \& Nesseris(2009)}]{Bogdanos:2009ib}
Bogdanos, C. \& Nesseris, S. 2009, JCAP, 05, 006

\bibitem[{Buchert(2008)}]{Buchert:2007ik}
Buchert, T. 2008, Gen. Rel. Grav., 40, 467

\bibitem[{Buchert {et~al.}(2015)}]{Buchert:2015iva}
Buchert, T. {et~al.} 2015, Class. Quant. Grav., 32, 215021

\bibitem[{Bull {et~al.}(2012)Bull, Clifton, \& Ferreira}]{Bull:2011wi}
Bull, P., Clifton, T., \& Ferreira, P.~G. 2012, Phys. Rev. D, 85, 024002

\bibitem[{Camarena {et~al.}(2021)Camarena, Marra, Sakr, \&
  Clarkson}]{Camarena:2021mjr}
Camarena, D., Marra, V., Sakr, Z., \& Clarkson, C. 2021, Mon. Not. Roy. Astron.
  Soc., 509, 1291

\bibitem[{C\'el\'erier(2000)}]{Celerier:1999hp}
C\'el\'erier, M.-N. 2000, Astron. Astrophys., 353, 63

\bibitem[{Chevallier \& Polarski(2001)}]{Chevallier:2000qy}
Chevallier, M. \& Polarski, D. 2001, Int. J. Mod. Phys. D, 10, 213

\bibitem[{Chiang {et~al.}(2019)Chiang, Romano, Nugier, \&
  Chen}]{Chiang:2017yrq}
Chiang, H.~W., Romano, A.~E., Nugier, F., \& Chen, P. 2019, JCAP, 11, 016

\bibitem[{Clarkson(2012)}]{Clarkson:2012bg}
Clarkson, C. 2012, Comptes Rendus Physique, 13, 682

\bibitem[{Clarkson {et~al.}(2008)Clarkson, Bassett, \& Lu}]{Clarkson:2007pz}
Clarkson, C., Bassett, B., \& Lu, T. H.-C. 2008, Phys. Rev. Lett., 101, 011301

\bibitem[{Clarkson {et~al.}(2011)Clarkson, Ellis, Larena, \&
  Umeh}]{Clarkson:2011zq}
Clarkson, C., Ellis, G., Larena, J., \& Umeh, O. 2011, Rept. Prog. Phys., 74,
  112901

\bibitem[{Clarkson \& Regis(2011)}]{Clarkson:2010ej}
Clarkson, C. \& Regis, M. 2011, JCAP, 02, 013

\bibitem[{{Conley} {et~al.}(2011){Conley}, {Guy}, {Sullivan}, {Regnault},
  {Astier}, {Balland}, {Basa}, {Carlberg}, {Fouchez}, {Hardin}, {Hook},
  {Howell}, {Pain}, {Palanque-Delabrouille}, {Perrett}, {Pritchet}, {Rich},
  {Ruhlmann-Kleider}, {Balam}, {Baumont}, {Ellis}, {Fabbro}, {Fakhouri},
  {Fourmanoit}, {Gonz{\'a}lez-Gait{\'a}n}, {Graham}, {Hudson}, {Hsiao},
  {Kronborg}, {Lidman}, {Mourao}, {Neill}, {Perlmutter}, {Ripoche}, {Suzuki},
  \& {Walker}}]{Conley:2011ku}
{Conley}, A., {Guy}, J., {Sullivan}, M., {et~al.} 2011, \apjs, 192, 1

\bibitem[{Costille {et~al.}(2018)Costille, Caillat, Rossin, Pascal, Sanchez,
  Barette, Laurent, Foulon, \& Pariès}]{NISP_paper}
Costille, A., Caillat, A., Rossin, C., {et~al.} 2018, in Space Telescopes and
  Instrumentation 2018: Optical, Infrared, and Millimeter Wave, ed. M.~Lystrup,
  H.~A. MacEwen, G.~G. Fazio, N.~Batalha, N.~Siegler, \& E.~C. Tong, Vol.
  10698, International Society for Optics and Photonics (SPIE), 730 -- 744

\bibitem[{Cropper {et~al.}(2018)Cropper, Pottinger, Azzollini, Szafraniec,
  Awan, Mellier, Berthé, Martignac, Cara, Giorgio, Sciortino, Bozzo, Genolet,
  Philippon, Hailey, Hunt, Swindells, Holland, Gow, Murray, Hall, Skottfelt,
  Amiaux, Laureijs, Racca, Salvignol, Short, Alvarez, Kitching, Hoekstra,
  Galli, Willis, Hu, Candini, Boucher, Bahlawan, Chaudery, de~Lacy, Pendem,
  Smit, Dubois, Horeau, Carty, Fontignie, Doumayrou, Larcheveque, Castelli,
  Cole, Niemi, Denniston, Massey, Kohley, Ferrando, \& Conversi}]{VIS_paper}
Cropper, M., Pottinger, S., Azzollini, R., {et~al.} 2018, in Space Telescopes
  and Instrumentation 2018: Optical, Infrared, and Millimeter Wave, ed.
  M.~Lystrup, H.~A. MacEwen, G.~G. Fazio, N.~Batalha, N.~Siegler, \& E.~C.
  Tong, Vol. 10698, International Society for Optics and Photonics (SPIE), 709
  -- 729

\bibitem[{de~la Torre \& Guzzo(2012)}]{delaTorre:2012dg}
de~la Torre, S. \& Guzzo, L. 2012, Mon. Not. Roy. Astron. Soc., 427, 327

\bibitem[{{DESI Collaboration: Aghamousa} {et~al.}(2016){DESI Collaboration:
  Aghamousa}, {Aguilar}, {Ahlen}, {Alam}, {Allen}, {Allende Prieto}, {Annis},
  {Bailey}, {Balland}, {Ballester}, {Baltay}, {Beaufore}, {Bebek}, {Beers},
  {Bell}, {Bernal}, {Besuner}, {Beutler}, {Blake}, {Bleuler}, {Blomqvist},
  {Blum}, {Bolton}, {Briceno}, {Brooks}, {Brownstein}, {Buckley-Geer},
  {Burden}, {Burtin}, {Busca}, {Cahn}, {Cai}, {Cardiel-Sas}, {Carlberg},
  {Carton}, {Casas}, {Castand er}, {Cervantes-Cota}, {Claybaugh}, {Close},
  {Coker}, {Cole}, {Comparat}, {Cooper}, {Cousinou}, {Crocce}, {Cuby},
  {Cunningham}, {Davis}, {Dawson}, {de la Macorra}, {De Vicente}, {Delubac},
  {Derwent}, {Dey}, {Dhungana}, {Ding}, {Doel}, {Duan}, {Ealet}, {Edelstein},
  {Eftekharzadeh}, {Eisenstein}, {Elliott}, {Escoffier}, {Evatt}, {Fagrelius},
  {Fan}, {Fanning}, {Farahi}, {Farihi}, {Favole}, {Feng}, {Fernandez},
  {Findlay}, {Finkbeiner}, {Fitzpatrick}, {Flaugher}, {Flender}, {Font-Ribera},
  {Forero-Romero}, {Fosalba}, {Frenk}, {Fumagalli}, {Gaensicke}, {Gallo},
  {Garc\'ia-Bellido}, {Gaztanaga}, {Pietro Gentile Fusillo}, {Gerard},
  {Gershkovich}, {Giannantonio}, {Gillet}, {Gonzalez-de-Rivera},
  {Gonzalez-Perez}, {Gott}, {Graur}, {Gutierrez}, {Guy}, {Habib}, {Heetderks},
  {Heetderks}, {Heitmann}, {Hellwing}, {Herrera}, {Ho}, {Holland}, {Honscheid},
  {Huff}, {Hutchinson}, {Huterer}, {Hwang}, {Illa Laguna}, {Ishikawa},
  {Jacobs}, {Jeffrey}, {Jelinsky}, {Jennings}, {Jiang}, {Jimenez}, {Johnson},
  {Joyce}, {Jullo}, {Juneau}, {Kama}, {Karcher}, {Karkar}, {Kehoe}, {Kennamer},
  {Kent}, {Kilbinger}, {Kim}, {Kirkby}, {Kisner}, {Kitanidis}, {Kneib},
  {Koposov}, {Kovacs}, {Koyama}, {Kremin}, {Kron}, {Kronig}, {Kueter-Young},
  {Lacey}, {Lafever}, {Lahav}, {Lambert}, {Lampton}, {Land riau}, {Lang},
  {Lauer}, {Le Goff}, {Le Guillou}, {Le Van Suu}, {Lee}, {Lee}, {Leitner},
  {Lesser}, {Levi}, {L'Huillier}, {Li}, {Liang}, {Lin}, {Linder}, {Loebman},
  {Luki{\'c}}, {Ma}, {MacCrann}, {Magneville}, {Makarem}, {Manera}, {Manser},
  {Marshall}, {Martini}, {Massey}, {Matheson}, {McCauley}, {McDonald},
  {McGreer}, {Meisner}, {Metcalfe}, {Miller}, {Miquel}, {Moustakas}, {Myers},
  {Naik}, {Newman}, {Nichol}, {Nicola}, {Nicolati da Costa}, {Nie}, {Niz},
  {Norberg}, {Nord}, {Norman}, {Nugent}, {O'Brien}, {Oh}, {Olsen}, {Padilla},
  {Padmanabhan}, {Padmanabhan}, {Palanque-Delabrouille}, {Palmese},
  {Pappalardo}, {P{\^a}ris}, {Park}, {Patej}, {Peacock}, {Peiris}, {Peng},
  {Percival}, {Perruchot}, {Pieri}, {Pogge}, {Pollack}, {Poppett}, {Prada},
  {Prakash}, {Probst}, {Rabinowitz}, {Raichoor}, {Ree}, {Refregier}, {Regal},
  {Reid}, {Reil}, {Rezaie}, {Rockosi}, {Roe}, {Ronayette}, {Roodman}, {Ross},
  {Ross}, {Rossi}, {Rozo}, {Ruhlmann-Kleider}, {Rykoff}, {Sabiu}, {Samushia},
  {Sanchez}, {Sanchez}, {Schlegel}, {Schneider}, {Schubnell}, {Secroun},
  {Seljak}, {Seo}, {Serrano}, {Shafieloo}, {Shan}, {Sharples}, {Sholl},
  {Shourt}, {Silber}, {Silva}, {Sirk}, {Slosar}, {Smith}, {Smoot}, {Som},
  {Song}, {Sprayberry}, {Staten}, {Stefanik}, {Tarle}, {Sien Tie}, {Tinker},
  {Tojeiro}, {Valdes}, {Valenzuela}, {Valluri}, {Vargas-Magana}, {Verde},
  {Walker}, {Wang}, {Wang}, {Weaver}, {Weaverdyck}, {Wechsler}, {Weinberg},
  {White}, {Yang}, {Yeche}, {Zhang}, {Zhao}, {Zheng}, {Zhou}, {Zhou}, {Zhu},
  {Zou}, \& {Zu}}]{DESI2016}
{DESI Collaboration: Aghamousa}, A., {Aguilar}, J., {Ahlen}, S., {et~al.} 2016,
  arXiv e-prints, arXiv:1611.00036

\bibitem[{Di~Valentino {et~al.}(2021)Di~Valentino, Mena, Pan, Visinelli, Yang,
  Melchiorri, Mota, Riess, \& Silk}]{DiValentino:2021izs}
Di~Valentino, E., Mena, O., Pan, S., {et~al.} 2021, Class. Quant. Grav., 38,
  153001

\bibitem[{Einstein \& Straus(1945)}]{Einstein:1945id}
Einstein, A. \& Straus, E.~G. 1945, Rev. Mod. Phys., 17, 120

\bibitem[{Eisenstein \& Hu(1998)}]{Eisenstein:1997ik}
Eisenstein, D.~J. \& Hu, W. 1998, \apj, 496, 605

\bibitem[{Ellis \& MacCallum(1969)}]{Ellis:1968vb}
Ellis, G. F.~R. \& MacCallum, M. A.~H. 1969, Commun. Math. Phys., 12, 108

\bibitem[{Ellis \& van Elst(1999)}]{Ellis:1998ct}
Ellis, G. F.~R. \& van Elst, H. 1999, NATO Sci. Ser. C, 541, 1

\bibitem[{{Euclid Collaboration: Paykari} {et~al.}(2020){Euclid Collaboration:
  Paykari}, {Kitching}, {Hoekstra}, {Azzollini}, {Cardone}, {Cropper},
  {Duncan}, {Kannawadi}, {Miller}, {Aussel}, {Conti}, {Auricchio}, {Baldi},
  {Bardelli}, {Biviano}, {Bonino}, {Borsato}, {Bozzo}, {Branchini},
  {Brau-Nogue}, {Brescia}, {Brinchmann}, {Burigana}, {Camera}, {Capobianco},
  {Carbone}, {Carretero}, {Castand er}, {Castellano}, {Cavuoti}, {Charles},
  {Cledassou}, {Colodro-Conde}, {Congedo}, {Conselice}, {Conversi}, {Copin},
  {Coupon}, {Courtois}, {Da Silva}, {Dupac}, {Fabbian}, {Farrens}, {Ferreira},
  {Fosalba}, {Fourmanoit}, {Frailis}, {Fumana}, {Galeotta}, {Garilli},
  {Gillard}, {Gillis}, {Giocoli}, {Graci{\'a}-Carpio}, {Grupp}, {Hormuth},
  {Ili{\'c}}, {Israel}, {Jahnke}, {Keihanen}, {Kermiche}, {Kilbinger},
  {Kirkpatrick}, {Kubik}, {Kunz}, {Kurki-Suonio}, {Laureijs}, {Le Mignant},
  {Ligori}, {Lilje}, {Lloro}, {Maciaszek}, {Maiorano}, {Marggraf}, {Markovic},
  {Martinet}, {Marulli}, {Massey}, {Mauri}, {Medinaceli}, {Mei}, {Mellier},
  {Meneghetti}, {Metcalf}, {Moresco}, {Moscardini}, {Munari}, {Neissner},
  {Nichol}, {Niemi}, {Nutma}, {Padilla}, {Paltani}, {Pasian}, {Pettorino},
  {Pires}, {Polenta}, {Raison}, {Renzi}, {Rhodes}, {Romelli}, {Roncarelli},
  {Rossetti}, {Saglia}, {Sakr}, {S{\'a}nchez}, {Sapone}, {Scaramella},
  {Schneider}, {Schrabback}, {Scottez}, {Secroun}, {Serrano}, {Sirignano},
  {Sirri}, {Stanco}, {Starck}, {Sureau}, {Tallada-Cresp{\'\i}}, {Taylor},
  {Tenti}, {Tereno}, {Toledo-Moreo}, {Torradeflot}, {Valenziano}, {Vannier},
  {Vassallo}, {Zoubian}, \& {Zucca}}]{Paykari2020}
{Euclid Collaboration: Paykari}, P., {Kitching}, T., {Hoekstra}, H., {et~al.}
  2020, \aap, 635, A139

\bibitem[{February {et~al.}(2010)February, Larena, Smith, \&
  Clarkson}]{February:2009pv}
February, S., Larena, J., Smith, M., \& Clarkson, C. 2010, Mon. Not. Roy.
  Astron. Soc., 405, 2231

\bibitem[{{Font-Ribera} {et~al.}(2014){Font-Ribera}, {McDonald}, {Mostek},
  {Reid}, {Seo}, \& {Slosar}}]{2014JCAP...05..023F}
{Font-Ribera}, A., {McDonald}, P., {Mostek}, N., {et~al.} 2014, JCAP, 05, 023

\bibitem[{Friedrich {et~al.}(2021)}]{Friedrich:2020dqo}
Friedrich, O. {et~al.} 2021, Mon. Not. Roy. Astron. Soc., 508, 3125

\bibitem[{Garc\'ia-Bellido \&
  Haugboelle(2008{\natexlab{a}})}]{GarciaBellido:2008nz}
Garc\'ia-Bellido, J. \& Haugboelle, T. 2008{\natexlab{a}}, JCAP, 0804, 003

\bibitem[{Garc\'ia-Bellido \&
  Haugboelle(2008{\natexlab{b}})}]{GarciaBellido:2008gd}
Garc\'ia-Bellido, J. \& Haugboelle, T. 2008{\natexlab{b}}, JCAP, 0809, 016

\bibitem[{Garc\'ia-Bellido \& Haugboelle(2009)}]{GarciaBellido:2008yq}
Garc\'ia-Bellido, J. \& Haugboelle, T. 2009, JCAP, 0909, 028

\bibitem[{Giblin {et~al.}(2016)Giblin, Mertens, \& Starkman}]{Giblin:2016mjp}
Giblin, J.~T., Mertens, J.~B., \& Starkman, G.~D. 2016, Astrophys. J., 833, 247

\bibitem[{Gong {et~al.}(2010)Gong, Cooray, \& Chen}]{Gong:2009yk}
Gong, Y., Cooray, A., \& Chen, X. 2010, \apj, 709, 1420

\bibitem[{Green \& Wald(2014)}]{Green:2014aga}
Green, S.~R. \& Wald, R.~M. 2014, Class. Quant. Grav., 31, 234003

\bibitem[{Harnois-Deraps {et~al.}(2019)Harnois-Deraps, Giblin, \&
  Joachimi}]{Harnois-Deraps:2019rsd}
Harnois-Deraps, J., Giblin, B., \& Joachimi, B. 2019, Astron. Astrophys., 631,
  A160

\bibitem[{Heavens {et~al.}(2011)Heavens, Jimenez, \& Maartens}]{Heavens:2011mr}
Heavens, A.~F., Jimenez, R., \& Maartens, R. 2011, JCAP, 09, 035

\bibitem[{Jonsson {et~al.}(2010)Jonsson, Sullivan, Hook, Basa, Carlberg,
  Conley, Fouchez, Howell, Perrett, \& Pritchet}]{Jonsson:2010wx}
Jonsson, J., Sullivan, M., Hook, I., {et~al.} 2010, Mon. Not. Roy. Astron.
  Soc., 405, 535

\bibitem[{Kaiser(2017)}]{Kaiser:2017hqn}
Kaiser, N. 2017, Mon. Not. Roy. Astron. Soc., 469, 744

\bibitem[{Kolb {et~al.}(2010)Kolb, Marra, \& Matarrese}]{Kolb:2009rp}
Kolb, E.~W., Marra, V., \& Matarrese, S. 2010, Gen.Rel.Grav., 42, 1399

\bibitem[{Kronborg {et~al.}(2010)}]{SNLS:2010rmd}
Kronborg, T. {et~al.} 2010, Astron. Astrophys., 514, A44

\bibitem[{{Laureijs} {et~al.}(2011){Laureijs}, {Amiaux}, {Arduini},
  {Augu{\`e}res}, {Brinchmann}, {Cole}, {Cropper}, {Dabin}, {Duvet}, {Ealet},
  {Garilli}, {Gondoin}, {Guzzo}, {Hoar}, {Hoekstra}, {Holmes}, {Kitching},
  {Maciaszek}, {Mellier}, {Pasian}, {Percival}, {Rhodes}, {Saavedra Criado},
  {Sauvage}, {Scaramella}, {Valenziano}, {Warren}, {Bender}, {Castander},
  {Cimatti}, {Le F{\`e}vre}, {Kurki-Suonio}, {Levi}, {Lilje}, {Meylan},
  {Nichol}, {Pedersen}, {Popa}, {Rebolo Lopez}, {Rix}, {Rottgering},
  {Zeilinger}, {Grupp}, {Hudelot}, {Massey}, {Meneghetti}, {Miller}, {Paltani},
  {Paulin-Henriksson}, {Pires}, {Saxton}, {Schrabback}, {Seidel}, {Walsh},
  {Aghanim}, {Amendola}, {Bartlett}, {Baccigalupi}, {Beaulieu}, {Benabed},
  {Cuby}, {Elbaz}, {Fosalba}, {Gavazzi}, {Helmi}, {Hook}, {Irwin}, {Kneib},
  {Kunz}, {Mannucci}, {Moscardini}, {Tao}, {Teyssier}, {Weller}, {Zamorani},
  {Zapatero Osorio}, {Boulade}, {Foumond}, {Di Giorgio}, {Guttridge}, {James},
  {Kemp}, {Martignac}, {Spencer}, {Walton}, {Bl{\"u}mchen}, {Bonoli},
  {Bortoletto}, {Cerna}, {Corcione}, {Fabron}, {Jahnke}, {Ligori}, {Madrid},
  {Martin}, {Morgante}, {Pamplona}, {Prieto}, {Riva}, {Toledo}, {Trifoglio},
  {Zerbi}, {Abdalla}, {Douspis}, {Grenet}, {Borgani}, {Bouwens}, {Courbin},
  {Delouis}, {Dubath}, {Fontana}, {Frailis}, {Grazian}, {Koppenh{\"o}fer},
  {Mansutti}, {Melchior}, {Mignoli}, {Mohr}, {Neissner}, {Noddle}, {Poncet},
  {Scodeggio}, {Serrano}, {Shane}, {Starck}, {Surace}, {Taylor},
  {Verdoes-Kleijn}, {Vuerli}, {Williams}, {Zacchei}, {Altieri}, {Escudero
  Sanz}, {Kohley}, {Oosterbroek}, {Astier}, {Bacon}, {Bardelli}, {Baugh},
  {Bellagamba}, {Benoist}, {Bianchi}, {Biviano}, {Branchini}, {Carbone},
  {Cardone}, {Clements}, {Colombi}, {Conselice}, {Cresci}, {Deacon}, {Dunlop},
  {Fedeli}, {Fontanot}, {Franzetti}, {Giocoli}, {Garc\'ia-Bellido}, {Gow},
  {Heavens}, {Hewett}, {Heymans}, {Holland}, {Huang}, {Ilbert}, {Joachimi},
  {Jennins}, {Kerins}, {Kiessling}, {Kirk}, {Kotak}, {Krause}, {Lahav}, {van
  Leeuwen}, {Lesgourgues}, {Lombardi}, {Magliocchetti}, {Maguire}, {Majerotto},
  {Maoli}, {Marulli}, {Maurogordato}, {McCracken}, {McLure}, {Melchiorri},
  {Merson}, {Moresco}, {Nonino}, {Norberg}, {Peacock}, {Pello}, {Penny},
  {Pettorino}, {Di Porto}, {Pozzetti}, {Quercellini}, {Radovich}, {Rassat},
  {Roche}, {Ronayette}, {Rossetti}, {Sartoris}, {Schneider}, {Semboloni},
  {Serjeant}, {Simpson}, {Skordis}, {Smadja}, {Smartt}, {Spano}, {Spiro},
  {Sullivan}, {Tilquin}, {Trotta}, {Verde}, {Wang}, {Williger}, {Zhao},
  {Zoubian}, \& {Zucca}}]{Laureijs:2011gra}
{Laureijs}, R., {Amiaux}, J., {Arduini}, S., {et~al.} 2011, arXiv e-prints,
  arXiv:1110.3193

\bibitem[{Laurent {et~al.}(2016)}]{Laurent:2016eqo}
Laurent, P. {et~al.} 2016, JCAP, 11, 060

\bibitem[{Lewis(2013)}]{Lewis:2013hha}
Lewis, A. 2013, Phys. Rev. D, 87, 103529

\bibitem[{Lewis \& Bridle(2002)}]{Lewis:2002ah}
Lewis, A. \& Bridle, S. 2002, Phys. Rev. D, 66, 103511

\bibitem[{Linder(2003)}]{Linder:2002et}
Linder, E.~V. 2003, PRL, 90, 091301

\bibitem[{{LSST Science Collaboration: Abell} {et~al.}(2009){LSST Science
  Collaboration: Abell}, {Allison}, {Anderson}, {Andrew}, {Angel}, {Armus},
  {Arnett}, {Asztalos}, {Axelrod}, {Bailey}, {Ballantyne}, {Bankert},
  {Barkhouse}, {Barr}, {Barrientos}, {Barth}, {Bartlett}, {Becker}, {Becla},
  {Beers}, {Bernstein}, {Biswas}, {Blanton}, {Bloom}, {Bochanski}, {Boeshaar},
  {Borne}, {Bradac}, {Brandt}, {Bridge}, {Brown}, {Brunner}, {Bullock},
  {Burgasser}, {Burge}, {Burke}, {Cargile}, {Chand rasekharan}, {Chartas},
  {Chesley}, {Chu}, {Cinabro}, {Claire}, {Claver}, {Clowe}, {Connolly}, {Cook},
  {Cooke}, {Cooray}, {Covey}, {Culliton}, {de Jong}, {de Vries}, {Debattista},
  {Delgado}, {Dell'Antonio}, {Dhital}, {Di Stefano}, {Dickinson}, {Dilday},
  {Djorgovski}, {Dobler}, {Donalek}, {Dubois-Felsmann}, {Durech},
  {Eliasdottir}, {Eracleous}, {Eyer}, {Falco}, {Fan}, {Fassnacht}, {Ferguson},
  {Fernandez}, {Fields}, {Finkbeiner}, {Figueroa}, {Fox}, {Francke}, {Frank},
  {Frieman}, {Fromenteau}, {Furqan}, {Galaz}, {Gal-Yam}, {Garnavich},
  {Gawiser}, {Geary}, {Gee}, {Gibson}, {Gilmore}, {Grace}, {Green}, {Gressler},
  {Grillmair}, {Habib}, {Haggerty}, {Hamuy}, {Harris}, {Hawley}, {Heavens},
  {Hebb}, {Henry}, {Hileman}, {Hilton}, {Hoadley}, {Holberg}, {Holman},
  {Howell}, {Infante}, {Ivezic}, {Jacoby}, {Jain}, {R}, {Jedicke}, {Jee},
  {Garrett Jernigan}, {Jha}, {Johnston}, {Jones}, {Juric}, {Kaasalainen},
  {Styliani}, {Kafka}, {Kahn}, {Kaib}, {Kalirai}, {Kantor}, {Kasliwal},
  {Keeton}, {Kessler}, {Knezevic}, {Kowalski}, {Krabbendam}, {Krughoff},
  {Kulkarni}, {Kuhlman}, {Lacy}, {Lepine}, {Liang}, {Lien}, {Lira}, {Long},
  {Lorenz}, {Lotz}, {Lupton}, {Lutz}, {Macri}, {Mahabal}, {Mandelbaum},
  {Marshall}, {May}, {McGehee}, {Meadows}, {Meert}, {Milani}, {Miller},
  {Miller}, {Mills}, {Minniti}, {Monet}, {Mukadam}, {Nakar}, {Neill}, {Newman},
  {Nikolaev}, {Nordby}, {O'Connor}, {Oguri}, {Oliver}, {Olivier}, {Olsen},
  {Olsen}, {Olszewski}, {Oluseyi}, {Padilla}, {Parker}, {Pepper}, {Peterson},
  {Petry}, {Pinto}, {Pizagno}, {Popescu}, {Prsa}, {Radcka}, {Raddick},
  {Rasmussen}, {Rau}, {Rho}, {Rhoads}, {Richards}, {Ridgway}, {Robertson},
  {Roskar}, {Saha}, {Sarajedini}, {Scannapieco}, {Schalk}, {Schindler},
  {Schmidt}, {Schmidt}, {Schneider}, {Schumacher}, {Scranton}, {Sebag},
  {Seppala}, {Shemmer}, {Simon}, {Sivertz}, {Smith}, {Allyn Smith}, {Smith},
  {Spitz}, {Stanford}, {Stassun}, {Strader}, {Strauss}, {Stubbs}, {Sweeney},
  {Szalay}, {Szkody}, {Takada}, {Thorman}, {Trilling}, {Trimble}, {Tyson}, {Van
  Berg}, {Vand en Berk}, {VanderPlas}, {Verde}, {Vrsnak}, {Walkowicz}, {Wand
  elt}, {Wang}, {Wang}, {Warner}, {Wechsler}, {West}, {Wiecha}, {Williams},
  {Willman}, {Wittman}, {Wolff}, {Wood-Vasey}, {Wozniak}, {Young}, {Zentner},
  \& {Zhan}}]{Abell:2009aa}
{LSST Science Collaboration: Abell}, P.~A., {Allison}, J., {Anderson}, S.~F.,
  {et~al.} 2009, arXiv e-prints, arXiv:0912.0201

\bibitem[{Marra {et~al.}(2007)Marra, Kolb, Matarrese, \& Riotto}]{Marra:2007pm}
Marra, V., Kolb, E.~W., Matarrese, S., \& Riotto, A. 2007, Phys.Rev., D76,
  123004

\bibitem[{Marra \& Notari(2011)}]{Marra:2011ct}
Marra, V. \& Notari, A. 2011, Class.Quant.Grav., 28, 164004

\bibitem[{Marra {et~al.}(2013)Marra, Quartin, \& Amendola}]{Amendola:2013twa}
Marra, V., Quartin, M., \& Amendola, L. 2013, Phys.Rev., D88, 063004

\bibitem[{Marra \& Sapone(2018)}]{Marra:2017pst}
Marra, V. \& Sapone, D. 2018, Phys. Rev. D, 97, 083510

\bibitem[{Martinelli {et~al.}(2020)}]{Martinelli:2020hud}
Martinelli, M. {et~al.} 2020, Astron. Astrophys., 644, A80

\bibitem[{Moffat \& Tatarski(1995)}]{Moffat:1994qy}
Moffat, J.~W. \& Tatarski, D.~C. 1995, Astrophys. J., 453, 17

\bibitem[{Mustapha {et~al.}(1997)Mustapha, Hellaby, \& Ellis}]{Mustapha:1998jb}
Mustapha, N., Hellaby, C., \& Ellis, G. F.~R. 1997, Mon. Not. Roy. Astron.
  Soc., 292, 817

\bibitem[{Nadolny {et~al.}(2021)Nadolny, Durrer, Kunz, \&
  Padmanabhan}]{Nadolny:2021hti}
Nadolny, T., Durrer, R., Kunz, M., \& Padmanabhan, H. 2021, JCAP, 11, 009

\bibitem[{Naselsky {et~al.}(2012)Naselsky, Zhao, Kim, \&
  Chen}]{Naselsky:2011jp}
Naselsky, P., Zhao, W., Kim, J., \& Chen, S. 2012, Astrophys. J., 749, 31

\bibitem[{Nesseris \& Garc\'ia-Bellido(2012)}]{Nesseris:2012tt}
Nesseris, S. \& Garc\'ia-Bellido, J. 2012, JCAP, 11, 033

\bibitem[{Nesseris \& Garc\'ia-Bellido(2013)}]{Nesseris:2013bia}
Nesseris, S. \& Garc\'ia-Bellido, J. 2013, PRD, 88, 063521

\bibitem[{Nesseris \& Sapone(2014)}]{Nesseris:2014vra}
Nesseris, S. \& Sapone, D. 2014, Phys. Rev. D, 90, 063006

\bibitem[{Nesseris \& Sapone(2015)}]{Nesseris:2014mfa}
Nesseris, S. \& Sapone, D. 2015, Int. J. Mod. Phys. D, 24, 1550045

\bibitem[{Nesseris {et~al.}(2015)Nesseris, Sapone, \&
  Garc\'\i{}a-Bellido}]{Nesseris:2014qca}
Nesseris, S., Sapone, D., \& Garc\'\i{}a-Bellido, J. 2015, Phys. Rev. D, 91,
  023004

\bibitem[{Nesseris \& Shafieloo(2010)}]{Nesseris:2010ep}
Nesseris, S. \& Shafieloo, A. 2010, MNRAS, 408, 1879

\bibitem[{{Peebles}(2020)}]{2020coce.book.....P}
{Peebles}, P.~J.~E. 2020, {Cosmology's Century: An Inside History of our Modern
  Understanding of the Universe}

\bibitem[{{Perivolaropoulos} \& {Skara}(2021)}]{Perivolaropoulos:2021jda}
{Perivolaropoulos}, L. \& {Skara}, F. 2021, arXiv e-prints, arXiv:2105.05208

\bibitem[{Pitrou {et~al.}(2021)Pitrou, Coc, Uzan, \& Vangioni}]{Pitrou:2020etk}
Pitrou, C., Coc, A., Uzan, J.-P., \& Vangioni, E. 2021, Mon. Not. Roy. Astron.
  Soc., 502, 2474

\bibitem[{{Planck Collaboration} {et~al.}(2020){Planck Collaboration},
  {Aghanim}, {Akrami}, {Ashdown}, {Aumont}, {Baccigalupi}, {Ballardini},
  {Banday}, {Barreiro}, {Bartolo}, {Basak}, {Battye}, {Benabed}, {Bernard},
  {Bersanelli}, {Bielewicz}, {Bock}, {Bond}, {Borrill}, {Bouchet}, {Boulanger},
  {Bucher}, {Burigana}, {Butler}, {Calabrese}, {Cardoso}, {Carron},
  {Challinor}, {Chiang}, {Chluba}, {Colombo}, {Combet}, {Contreras}, {Crill},
  {Cuttaia}, {de Bernardis}, {de Zotti}, {Delabrouille}, {Delouis}, {Di
  Valentino}, {Diego}, {Dor{\'e}}, {Douspis}, {Ducout}, {Dupac}, {Dusini},
  {Efstathiou}, {Elsner}, {En{\ss}lin}, {Eriksen}, {Fantaye}, {Farhang},
  {Fergusson}, {Fernandez-Cobos}, {Finelli}, {Forastieri}, {Frailis},
  {Fraisse}, {Franceschi}, {Frolov}, {Galeotta}, {Galli}, {Ganga},
  {G{\'e}nova-Santos}, {Gerbino}, {Ghosh}, {Gonz{\'a}lez-Nuevo}, {G{\'o}rski},
  {Gratton}, {Gruppuso}, {Gudmundsson}, {Hamann}, {Handley}, {Hansen},
  {Herranz}, {Hildebrandt}, {Hivon}, {Huang}, {Jaffe}, {Jones}, {Karakci},
  {Keih{\"a}nen}, {Keskitalo}, {Kiiveri}, {Kim}, {Kisner}, {Knox},
  {Krachmalnicoff}, {Kunz}, {Kurki-Suonio}, {Lagache}, {Lamarre}, {Lasenby},
  {Lattanzi}, {Lawrence}, {Le Jeune}, {Lemos}, {Lesgourgues}, {Levrier},
  {Lewis}, {Liguori}, {Lilje}, {Lilley}, {Lindholm}, {L{\'o}pez-Caniego},
  {Lubin}, {Ma}, {Mac{\'\i}as-P{\'e}rez}, {Maggio}, {Maino}, {Mandolesi},
  {Mangilli}, {Marcos-Caballero}, {Maris}, {Martin}, {Martinelli},
  {Mart{\'\i}nez-Gonz{\'a}lez}, {Matarrese}, {Mauri}, {McEwen}, {Meinhold},
  {Melchiorri}, {Mennella}, {Migliaccio}, {Millea}, {Mitra},
  {Miville-Desch{\^e}nes}, {Molinari}, {Montier}, {Morgante}, {Moss}, {Natoli},
  {N{\o}rgaard-Nielsen}, {Pagano}, {Paoletti}, {Partridge}, {Patanchon},
  {Peiris}, {Perrotta}, {Pettorino}, {Piacentini}, {Polastri}, {Polenta},
  {Puget}, {Rachen}, {Reinecke}, {Remazeilles}, {Renzi}, {Rocha}, {Rosset},
  {Roudier}, {Rubi{\~n}o-Mart{\'\i}n}, {Ruiz-Granados}, {Salvati}, {Sandri},
  {Savelainen}, {Scott}, {Shellard}, {Sirignano}, {Sirri}, {Spencer},
  {Sunyaev}, {Suur-Uski}, {Tauber}, {Tavagnacco}, {Tenti}, {Toffolatti},
  {Tomasi}, {Trombetti}, {Valenziano}, {Valiviita}, {Van Tent}, {Vibert},
  {Vielva}, {Villa}, {Vittorio}, {Wandelt}, {Wehus}, {White}, {White},
  {Zacchei}, \& {Zonca}}]{Aghanim:2018eyx}
{Planck Collaboration}, {Aghanim}, N., {Akrami}, Y., {et~al.} 2020, \aap, 641,
  A6

\bibitem[{Pozzetti {et~al.}(2016)Pozzetti, Hirata, Geach, Cimatti, Baugh,
  Cucciati, Merson, Norberg, \& Shi}]{Pozzetti:2016cch}
Pozzetti, L., Hirata, C.~M., Geach, J.~E., {et~al.} 2016, A\&A, 590, A3

\bibitem[{Quartin {et~al.}(2014)Quartin, Marra, \& Amendola}]{Quartin:2013moa}
Quartin, M., Marra, V., \& Amendola, L. 2014, Phys.Rev., D89, 023009

\bibitem[{Racca {et~al.}(2016)}]{Racca:2016qpi}
Racca, G.~D. {et~al.} 2016, Proc. SPIE Int. Soc. Opt. Eng., 9904, 0O

\bibitem[{Redlich {et~al.}(2014)Redlich, Bolejko, Meyer, Lewis, \&
  Bartelmann}]{Redlich:2014gga}
Redlich, M., Bolejko, K., Meyer, S., Lewis, G.~F., \& Bartelmann, M. 2014,
  Astron. Astrophys., 570, A63

\bibitem[{Reichardt {et~al.}(2021)}]{Reichardt:2020jrr}
Reichardt, C.~L. {et~al.} 2021, Astrophys. J., 908, 199

\bibitem[{Sahni {et~al.}(2008)Sahni, Shafieloo, \& Starobinsky}]{Sahni:2008xx}
Sahni, V., Shafieloo, A., \& Starobinsky, A.~A. 2008, Phys. Rev. D, 78, 103502

\bibitem[{Sakr {et~al.}(2018)Sakr, Ili\'c, Blanchard, Bittar, \&
  Farah}]{Sakr:2018new}
Sakr, Z., Ili\'c, S., Blanchard, A., Bittar, J., \& Farah, W. 2018, Astron.
  Astrophys., 620, A78

\bibitem[{Sapone {et~al.}(2014)Sapone, Majerotto, \& Nesseris}]{Sapone:2014nna}
Sapone, D., Majerotto, E., \& Nesseris, S. 2014, PRD, 90, 023012

\bibitem[{{Scolnic} {et~al.}(2018){Scolnic}, {Jones}, {Rest}, {Pan},
  {Chornock}, {Foley}, {Huber}, {Kessler}, {Narayan}, {Riess}, {Rodney},
  {Berger}, {Brout}, {Challis}, {Drout}, {Finkbeiner}, {Lunnan}, {Kirshner},
  {Sand ers}, {Schlafly}, {Smartt}, {Stubbs}, {Tonry}, {Wood-Vasey}, {Foley},
  {Hand}, {Johnson}, {Burgett}, {Chambers}, {Draper}, {Hodapp}, {Kaiser},
  {Kudritzki}, {Magnier}, {Metcalfe}, {Bresolin}, {Gall}, {Kotak}, {McCrum}, \&
  {Smith}}]{Scolnic:2017caz}
{Scolnic}, D.~M., {Jones}, D.~O., {Rest}, A., {et~al.} 2018, \apj, 859, 101

\bibitem[{Scrimgeour {et~al.}(2012)}]{Scrimgeour:2012wt}
Scrimgeour, M. {et~al.} 2012, Mon. Not. Roy. Astron. Soc., 425, 116

\bibitem[{Seikel {et~al.}(2012)Seikel, Yahya, Maartens, \&
  Clarkson}]{Seikel:2012cs}
Seikel, M., Yahya, S., Maartens, R., \& Clarkson, C. 2012, Phys. Rev. D, 86,
  083001

\bibitem[{Tomita(2000)}]{Tomita:1999qn}
Tomita, K. 2000, Astrophys. J., 529, 38

\bibitem[{Torrado \& Lewis(2021)}]{Torrado:2020dgo}
Torrado, J. \& Lewis, A. 2021, JCAP, 05, 057

\bibitem[{Valkenburg {et~al.}(2014)Valkenburg, Marra, \&
  Clarkson}]{Valkenburg:2012td}
Valkenburg, W., Marra, V., \& Clarkson, C. 2014, Mon. Not. Roy. Astron. Soc.,
  438, L6

\bibitem[{von Marttens {et~al.}(2019{\natexlab{a}})von Marttens, Casarini,
  Mota, \& Zimdahl}]{vonMarttens:2018iav}
von Marttens, R., Casarini, L., Mota, D.~F., \& Zimdahl, W. 2019{\natexlab{a}},
  Phys. Dark Univ., 23, 100248

\bibitem[{von Marttens {et~al.}(2021)von Marttens, Gonzalez, Alcaniz, Marra, \&
  Casarini}]{vonMarttens:2020apn}
von Marttens, R., Gonzalez, J.~E., Alcaniz, J., Marra, V., \& Casarini, L.
  2021, Phys. Rev. D, 104, 043515

\bibitem[{von Marttens {et~al.}(2019{\natexlab{b}})von Marttens, Marra,
  Casarini, Gonzalez, \& Alcaniz}]{vonMarttens:2018bvz}
von Marttens, R., Marra, V., Casarini, L., Gonzalez, J.~E., \& Alcaniz, J.
  2019{\natexlab{b}}, Phys. Rev. D, 99, 043521

\bibitem[{Wang {et~al.}(2013)Wang, Chuang, \& Hirata}]{Wang:2012bx}
Wang, Y., Chuang, C.-H., \& Hirata, C.~M. 2013, \mnras, 430, 2446

\bibitem[{Yahya {et~al.}(2014)Yahya, Seikel, Clarkson, Maartens, \&
  Smith}]{Yahya:2013xma}
Yahya, S., Seikel, M., Clarkson, C., Maartens, R., \& Smith, M. 2014, Phys.
  Rev. D, 89, 023503

\bibitem[{Zhang \& Stebbins(2011)}]{Zhang:2010fa}
Zhang, P. \& Stebbins, A. 2011, Phys.Rev.Lett., 107, 041301

\bibitem[{Zibin(2011)}]{Zibin:2011ma}
Zibin, J.~P. 2011, Phys.Rev., D84, 123508

\bibitem[{Zibin \& Moss(2011)}]{Moss:2011ze}
Zibin, J.~P. \& Moss, A. 2011, Class.Quant.Grav., 28, 164005

\bibitem[{Zunckel \& Clarkson(2008)}]{Zunckel:2008ti}
Zunckel, C. \& Clarkson, C. 2008, Phys. Rev. Lett., 101, 181301

\end{thebibliography}

\clearpage

\begin{appendix} 
\section{Improvement factors\label{app:quantimp}}
Here we briefly present our method for quantifying the improvement factors expected from the forthcoming surveys compared to the currently available data. We use the GA and two complimentary approaches to do this.

First, we calculate the average deviation of the GA best-fit in units of sigma values from the null hypothesis (either the best-fitting \lcdm or the fiducial \lcdm, depending on the data) over the whole redshift range of the data for each of the tests. Then we estimate the ratio of the average deviations for the current and mock data in order to quantify the improvement brought by Euclid. In particular, assuming the GA produces a reconstruction function for one of the null tests, denoted by $f_\mathrm{GA}(z)$ along with some error $\sigma_{f_\mathrm{GA}}(z)$, we can quantify the average absolute deviation in sigma values from the null hypothesis as
\be
\langle D_\mathrm{null} \rangle=\frac{1}{z_\mathrm{max}-z_\mathrm{min}}\,\int_{z_\mathrm{min}}^{z_\mathrm{max}}\,dz\, \left|\frac{f_\mathrm{GA}(z)-f_\mathrm{null}(z)}{\sigma_{f_\mathrm{GA}}(z)}\right|, \label{eq:dnull}
\ee
where $f_\mathrm{null}(z)$ is the value of the null test for the null hypothesis, and $z_\mathrm{min}$, $z_\mathrm{max}$ are the minimum and maximum values of the redshift range of the data.

Second, for consistency with previous analyses \citep[see][]{Martinelli:2020hud}, we also estimate the average size of the errors over the whole redshift range, which is similar to a figure of merit, thus providing an estimate by how much forthcoming surveys will improve the errors in the reconstructions of the test. Specifically, to quantify this improvement, we calculate the average size of the errors in the redshift range of the data as follows:
\be
\langle \sigma_\mathrm{GA} \rangle=\frac{1}{z_\mathrm{max}-z_\mathrm{min}}\,\int_{z_\mathrm{min}}^{z_\mathrm{max}}\,dz\, \sigma_{f_\mathrm{GA}}(z). \label{eq:averr}
\ee

\begin{table}
\begin{center}
\setlength{\tabcolsep}{4pt}
\renewcommand{\arraystretch}{1.35}
\caption{Improvement factors for the average absolute deviation in sigma values from the null hypothesis $\langle D_\mathrm{null} \rangle$ given by \Cref{eq:dnull} and the average size of the errors $\langle \sigma_\mathrm{GA} \rangle$ given by \Cref{eq:averr} in the redshift range of the current and mock data, along with the corresponding improvement ratio. In the last two rows, we also present the average and median improvement factor for all the tests. \label{tab:factors}}
\begin{tabular}{cccc}
\hline
\hline
Test & Data/ratio & $\langle D_\mathrm{null} \rangle$ & $\langle \sigma_\mathrm{GA} \rangle$ \\ \\
 \hline
    & Current & $0.081$ & $0.260$ \\
$\textrm{Om}_\mathrm{H}(z)$ & \Euclid + DESI + LSST & $0.181$ & $0.004$ \\
    & ratio & $0.450$ & $73.497$ \\
 \\
    & Current & $0.238$ & $0.156$ \\
$\Sigma(z)$ & \Euclid + DESI + LSST & $0.199$ & $0.048$ \\
    & ratio & $1.195$ & $3.218$ \\
 \\
    & Current & $0.093$ & $0.557$ \\
$r_0$ & \Euclid + DESI + LSST & $0.180$ & $0.007$ \\
    & ratio & $0.517$ & $76.338$ \\
 \\
    & Current & $0.422$ & $0.039$ \\
$\textrm{Om}_\textrm{dL}$ & \Euclid + DESI + LSST & $0.119$ & $0.080$ \\
    & ratio & $3.546$ & $0.492$ \\
 \\
    & Current & $0.422$ & $0.039$ \\
$\textrm{Om}_\textrm{dA}$ & \Euclid + DESI + LSST & $0.119$ & $0.080$ \\
    & ratio & $3.546$ & $0.492$ \\
 \\
    & Current & $0.251$ & $0.864$ \\
$z^2 \Omega_{\rm k}(z)$ & \Euclid + DESI + LSST & $0.126$ & $0.384$ \\
    & ratio & $1.990$ & $2.248$ \\
 \\
    & Current & $0.593$ & $0.384$ \\
$\mathcal{O}_{\rm K}(z)$ & \Euclid + DESI + LSST & $0.080$ & $0.098$ \\
    & ratio & $7.457$ & $3.918$ \\
 \\
    & Current & $0.450$ & $0.175$ \\
$\mathcal{O}_{\rm m}(z)$ & \Euclid + DESI + LSST & $0.087$ & $0.040$ \\
    & ratio & $5.147$ & $4.413$ \\
 \\
Summary & Average & $2.981$ & $20.577$ \\
        & Median  & $2.768$ & $3.568$ \\
\hline
\hline
\end{tabular}\\
\end{center}
\end{table}

In \Cref{tab:factors} we present the improvement factors for the average absolute deviation in sigma values from the null hypothesis $\langle D_\mathrm{null} \rangle$ given by \Cref{eq:dnull} and the average size of the errors $\langle \sigma_\mathrm{GA} \rangle$  given by \Cref{eq:averr} in the redshift range of the current and mock data, along with the corresponding improvement. In the last two rows, we also present the average and median improvement factor across all the tests. Specifically, as listed in \Cref{tab:factors}, data from forthcoming surveys will bring an improvement factor of about a factor of three or more for most null tests. In order to reduce the possible double counting, as some of the tests are strongly correlated with each other, we also consider the median of the ratios of the improvement factors of the tests, which also gives an improvement of about a factor of three. Hence, in order to be conservative, we quote a factor of three as our final improvement factor when we compare the reconstructions of the null tests using the GA for the current and forthcoming data.
\end{appendix}

\end{document}